\documentclass[english,american,pre, reprint, superscriptaddress, nofootinbib]{revtex4-1}
\usepackage[T1]{fontenc}
\usepackage[latin9]{inputenc}
\usepackage{color}
\usepackage{babel}
\usepackage{units}
\usepackage{amsmath}
\usepackage{amssymb}
\usepackage{graphicx}
\usepackage{esint}
\usepackage[unicode=true,
 bookmarks=true,bookmarksnumbered=false,bookmarksopen=false,
 breaklinks=false,pdfborder={0 0 1},backref=false,colorlinks=false]
 {hyperref}
\hypersetup{pdftitle={Fluctuations and information filtering in coupled populations of spiking neurons with adaptation},
 pdfauthor={Moritz Deger, Tilo Schwalger, Richard Naud, Wulfram Gerstner}}

\makeatletter

\providecommand{\tabularnewline}{\\}

\makeatother

\begin{document}

\title{Fluctuations and information filtering in coupled populations of
spiking neurons with adaptation}

\thanks{This article was published as: M.~Deger, T.~Schwalger, R.~Naud,
W.~Gerstner, Fluctuations and information filtering in coupled populations
of spiking neurons with adaptation, Phys. Rev. E 90, 062704 (2014),
\href{http://dx.doi.org/10.1103/PhysRevE.90.062704}{DOI: 10.1103/PhysRevE.90.062704}~.
\\
Typesetting errors were corrected in the paragraph of Eq.~(\ref{eq:def_theta_QR})
and in Eqs.~(\ref{eq:QR_kernel_average}) and (\ref{eq:coherence_iext}).}

\author{Moritz Deger}

\affiliation{School of Computer and Communication Sciences and School of Life
Sciences, Brain Mind Institute, \'Ecole polytechnique f\'ed\'erale
de Lausanne, Station 15, 1015 Lausanne EPFL, Switzerland}

\thanks{Corresponding authors: M.~D., moritz.deger@epfl.ch~; \\
T.~S., tilo.schwalger@epfl.ch~.}

\author{Tilo Schwalger}

\affiliation{School of Computer and Communication Sciences and School of Life
Sciences, Brain Mind Institute, \'Ecole polytechnique f\'ed\'erale
de Lausanne, Station 15, 1015 Lausanne EPFL, Switzerland}

\author{Richard Naud}

\affiliation{Department of Physics, University of Ottawa, 150 Louis Pasteur, K1N-6N5
Ottawa, Ontario, Canada}

\author{Wulfram Gerstner}

\affiliation{School of Computer and Communication Sciences and School of Life
Sciences, Brain Mind Institute, \'Ecole polytechnique f\'ed\'erale
de Lausanne, Station 15, 1015 Lausanne EPFL, Switzerland}

\date{\today}

\begin{abstract}
Finite-sized populations of spiking elements are fundamental to brain
function, but also used in many areas of physics. Here we present
a theory of the dynamics of finite-sized populations of spiking units,
based on a quasi-renewal description of neurons with adaptation. We
derive an integral equation with colored noise that governs the stochastic
dynamics of the population activity in response to time-dependent
stimulation and calculate the spectral density in the asynchronous
state. We show that systems of coupled populations with adaptation
can generate a frequency band in which sensory information is preferentially
encoded. The theory is applicable to fully as well as randomly connected
networks, and to leaky integrate-and-fire as well as to generalized
spiking neurons with adaptation on multiple time scales.
\end{abstract}
\maketitle

\section{Introduction }

Multi-scale modeling of complex systems has led to important advances
in fields as diverse as complex fluid dynamics, chemical biology,
soft matter physics, meteorology, computer science, and neuroscience
\citep{Klein2008,Ayton2007,Peter2009,Pielke2002,Benaim2008,Gerstner2012}.
In these approaches, mathematical methods such as mean-field theories
and coarse-graining provide the basis to link properties of microscopic
elements to macroscopic variables. In many cases, macroscopic variables
fluctuate due to a finite number of microscopic elements. For instance,
in the brain, neurons can be grouped into populations of $50$ to
$1000$ neurons \citep{Lefort2009} with similar properties \citep{Hubel1962,Mensi2012}.
Fluctuations of the global activity of such populations are not captured
in classical mean-field theories \citep{Brunel1999,Gerstner2000},
which assume infinite system size. Here we put forward a theory for
the fluctuating macroscopic activity in networks of pulse-coupled
elements occurring in neuronal networks \citep{Pillow2008}, queuing
theory \citep{Frost1994} and synchronizing fireflies \citep{Mirollo90}. 

Finite-size effects in networks of spiking elements have been approached
by different methods, including extensions of the Fokker-Planck equation
for neuronal membrane potentials \citep{Brunel1999,Mattia2002}, stochastic
field theory \citep{Buice2013}, moment expansions in networks of
generalized linear models (GLM) \citep{Toyoizumi2009}, modified Hawkes
processes \citep{Hawkes1971a,Pernice2012,Helias2013}, simplified
Markov neuron models \citep{Soula2007,Buice2007,Bressloff2009,Benayoun2010a,Touboul2011,Dumont2014,Lagzi2014}
and the use of a linear response formalism for spike trains perturbed
by finite-size fluctuations \citep{Lindner2005,Akerberg2009,Trousdale2012}.
These studies lack, however, slow cellular feedback mechanisms mediating
adaptation. 

Adaptation characterized by a reduced response of a neuron to slow
compared to fast inputs is a wide-spread phenomenon in the brain and
has important implications for signal processing \citep{Benda2003c,Lundstrom2008,Pozzorini2013,Farkhooi2013}
and the spontaneous activity of single neurons \citep{Prescott2008,Schwalger2013a,Schwalger2013}.
On the population level, adaptation has been recently analyzed using
a quasi-renewal (QR) theory \citep{Naud2012}. The QR framework uses
tools of renewal point process theory \citep{Cox1980} to treat neurons
with arbitrary refractoriness. In particular, the dynamics of the
population activity is determined by an integral equation \citep{Wilson1972,Gerstner2000,Deger2010}.
These studies, however, have assumed an infinitely large population.

Here we present a theory for the interaction of finite-sized populations
of adapting neurons. The theory is valid for the broad class of neuron
models that can be approximated by a QR point process. This includes
integrate-and-fire (IF) as well as GLM neurons, for which parameters
can be reliably extracted from experimental data \citep{Pillow2008,Truccolo2010a,Mensi2012}.

Based on this theory we analyze information filtering (noise shaping)
in neuronal populations. We show that in a single population no noise
shaping occurs, but that in coupled populations, band-pass-like noise
shaping is possible due to adaptation and connectivity. 

This article is structured as follows: First, we present the general
dynamics of the population activity and its fluctuations. We then
describe how randomly connected, adapting neurons can be treated in
this framework. For fluctuations about a stationary state, we linearize
the dynamics and compute the spectral density of the population activity.
We then determine its coherence with external input signals and quantify
information transmission. 
\begin{figure}[t]
\begin{centering}
\includegraphics[width=1\columnwidth]{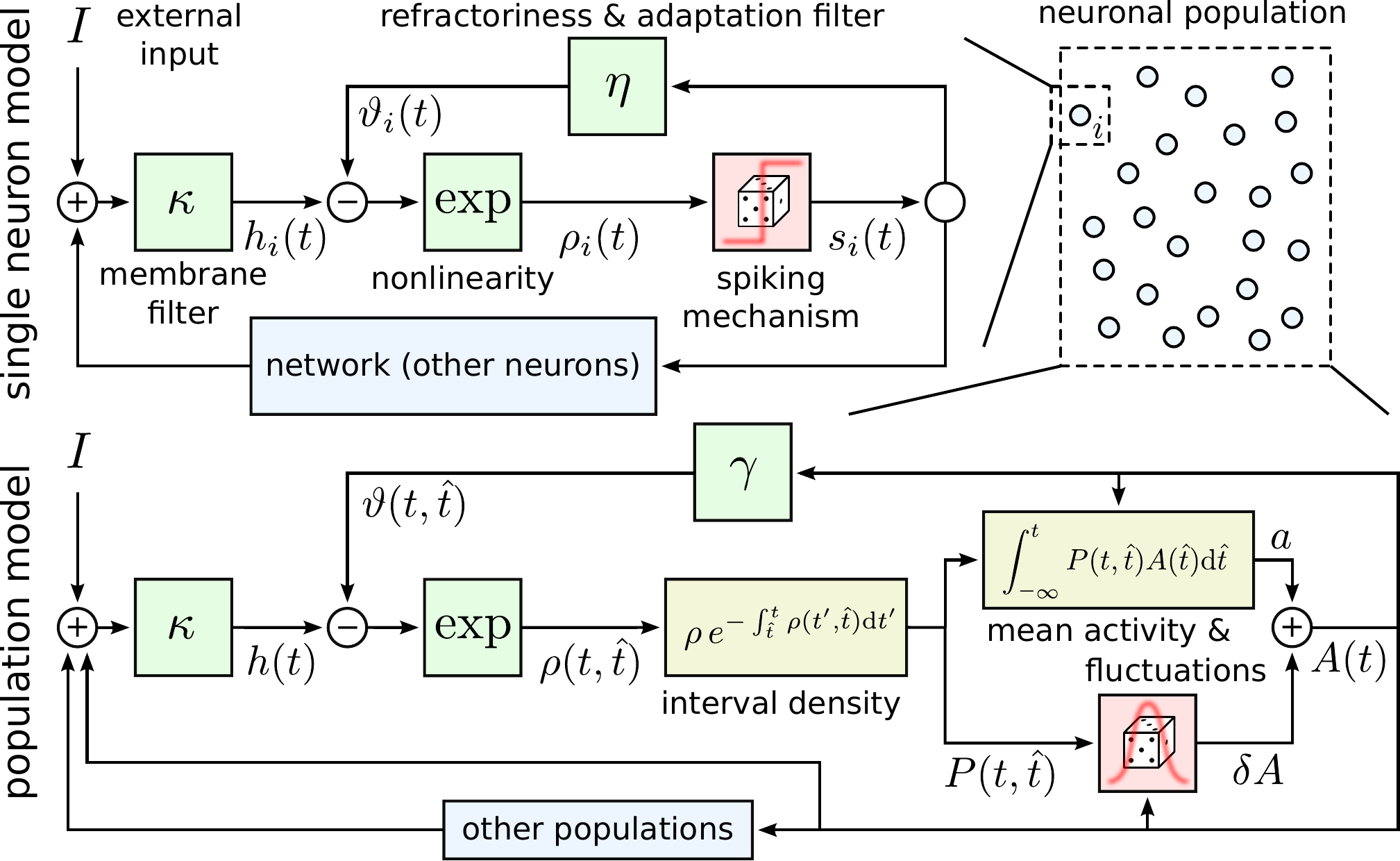}
\par\end{centering}

\caption{(Color) Schematic of the spike response model (a GLM for spike generation)
with exponential escape noise and the derived quasi-renewal population
model.\label{fig:schematics}}
\end{figure}

\section{Results}

\subsection{Dynamics of globally coupled renewal models.~}

Our main quantity of interest is the population activity $A(t)=N^{-1}\sum_{i=1}^{N}s_{i}(t),$
where $s_{i}(t)=\sum_{k}\delta(t-t_{i}^{k})$ is the spike train of
neuron $i$ with spike times $t_{i}^{k}$ and $N$ denotes the number
of neurons. In experiments or simulations, the measured activity $\bar{A}(t)$
would be determined by temporal filtering of the population activity,
i.e. $\bar{A}(t)=\int_{-\infty}^{\infty}A(t-s)f(s)\mathrm{d}s$ with
a normalized filter function $f(s)$ with finite support. Below we
will use a rectangular filter $f(s)=\theta(s)\theta(\Delta t-s)/\Delta t$,
where $\theta(s)$ is the Heaviside step function. 

To determine the fluctuation statistics of $A(t)$, we generalize
the integral equation of an infinite population \citep{Gerstner2000}
to large but finite $N$. Let us first consider a homogeneous population
of all-to-all connected renewal neurons. In this case, the spikes
of each neuron occur with an instantaneous rate or hazard function
$\rho_{\mathcal{H}}(t,\hat{t})$, which only depends on its last spike
time $\hat{t}\leq t$ and the synaptic input determined by the history
$\mathcal{H}(t)=\{A(t^{\prime})\}_{t^{\prime}<t}$ of the population
activity. Note that for uncoupled stationary networks, the hazard
reduces to $\rho_{\mathcal{H}}(t,\hat{t})=\rho(t-\hat{t})$ as it
should be for a renewal model. The probability density of the next
spike time $t$ given $\hat{t}$ is given by $P_{\mathcal{H}}(t,\hat{t})=\rho_{\mathcal{H}}(t,\hat{t})S_{\mathcal{H}}(t,\hat{t})$,
with the survivor function defined as $S_{\mathcal{H}}(t,\hat{t})=\exp(-\int_{\hat{t}}^{t}\rho_{\mathcal{H}}(t^{\prime},\hat{t})dt^{\prime})$. 

Our approach is to use the Gaussian approximation for large $N$,
i.e. we calculate the first- and second-order statistics of $A(t)$
as a functional of\emph{ }its past activity (see Appendix~\ref{sec:derivation-fn}).
However, the dynamics of $A$ depends on its own history $\mathcal{H}(t)$
and on the occupation density of refractory states $t-\hat{t}$ across
the population \citep{Meyer2002}. Thus, we have to average over
the possible refractory states consistent with a given history $\mathcal{H}$.
To perform the Gaussian approximation, the dynamics of the full system
is coarse-grained by discretizing time with a small time step $\Delta t$
which is still large enough to include many spikes of the population.
For large $N$, the number of neurons that fire in the time bin at
$t$ and had their last spike in bin $\hat{t}$ is a Gaussian random
number with mean and variance $n_{0}(\hat{t})P_{\mathcal{H}}(t,\hat{t})\Delta t$,
where $n_{0}(\hat{t})=N\int_{\hat{t}}^{\hat{t}+\Delta t}A(s)\, ds$
is the past spike count at $\hat{t}<t$. Summing over $\hat{t}$ and
treating $\Delta t$ as macroscopically infinitesimal, we find the
conditional mean activity (see Appendix~\ref{sec:derivation-fn})
\begin{align}
a(t) & =\int_{-\infty}^{t}P_{\mathcal{H}}(t,\hat{t})A(\hat{t})\,\mathrm{d}\hat{t},\label{eq:pop_int_abar}
\end{align}
which is equal to the population integral \citep{Gerstner2000} for
the infinite system. For finite $N$, $A(t)$ will be of the form
\begin{align}
A(t) & =a(t)+\delta A(t),\label{eq:pop_dyn}
\end{align}
where the deviation $\delta A(t)$ has zero mean and a diverging standard
deviation $\sqrt{a(t)/(N\Delta t)}$ because the variance of the spike
count in $[t,t+\Delta t]$ is given by $N\Delta t\, a(t)$. Importantly,
$\delta A(t)$ cannot be described by a white noise process but future
values $\delta A(t+\tau)$, $\tau>0$, are correlated with $\delta A(t)$,
because they share a common history $\mathcal{H}(t)$. In fact, a
neuron that fired its last spike at $\hat{t}<t$ cannot have its \textit{next}
spike at both times $t$ and $t+\tau$, which induces a negative correlation
for the deviations at $t$ and $t+\tau$. We find (see Appendix~\ref{sec:derivation-fn})
for $\tau\ge0$ the conditional correlation function 
\begin{multline}
\langle\delta A(t+\tau)\delta A(t)\rangle_{\mathcal{H}(t)}=N^{-1}a(t)\delta(\tau)-\\
N^{-1}\int_{-\infty}^{t}P_{\mathcal{H}}(t+\tau,\hat{t})P_{\mathcal{H}}(t,\hat{t})A(\hat{t})\,\mathrm{d}\hat{t},\label{eq:corr_xi}
\end{multline}
where $\langle\cdot\rangle_{\mathcal{H}(t)}$ denotes the average
conditioned on the history of $A$ before $t$. Thus, the correlation
function is in general explicitly time-dependent.

\subsection{Adaptation and random connectivity.~}

In the presence of adaptation, the instantaneous rate of a neuron
depends on all its previous spikes so that it can no longer be described
by renewal theory. Here we describe how adapting neurons in networks
may still be approximated by a quasi-renewal process. Specifically,
we consider a homogeneous population of neurons modeled by the spike-response
model with escape-noise \citep{Gerstner2000}, also known as GLM \citep{Pillow2008,Truccolo2010a,Mensi2012},
with hazard function 
\begin{align}
\rho_{i}(t) & =c\exp\left[(h_{i}(t)-\vartheta_{i}(t))/\delta u\right],\label{eq:rho_esc_noise_full}
\end{align}
(Fig.~\ref{fig:schematics}). That is, neuron $i$ produces a spike
in a small time interval $[t,\Delta t)$ with probability $\rho_{i}(t)\Delta t$.
This probability depends on the input potential $h_{i}(t)=\left[\kappa\ast\left(\sum_{j=1}^{N}w_{ij}s_{j}(t)+I\right)\right](t)$,
which is driven by presynaptic spike trains $s_{j}(t)$ (with synaptic
weight $w_{ij}$) and external input $I(t)$. The membrane filter
kernel is given by $\kappa(t)=\theta(t-\tau_{\mathrm{s}})\exp(-(t-\tau_{\mathrm{s}})/\tau_{\mathrm{m}})$,
where $\tau_{\mathrm{s}}$ and $\tau_{\mathrm{m}}$ are the synaptic
delay and the membrane time constant, respectively. The operation
$\ast$ denotes the convolution $(f\ast g)(t)=\int_{-\infty}^{\infty}f(t-s)g(s)\mathrm{d}s$
and $\theta(t)$ is the Heaviside step function. The variable $\vartheta_{i}$
defined as $\vartheta_{i}(t)=(s_{i}\ast\eta)(t)$ can be interpreted
as a dynamic firing threshold that is triggered by the neuron's own
output spike train $s_{i}(t)$ \citep{Mensi2012}. Here, $\eta(t)$
is a feedback kernel that consists of two parts, $\eta=\eta_{\mathrm{a}}+\eta_{\mathrm{r}}$:
a short-range refractory kernel $\eta_{\mathrm{r}}(t)=\theta(t)\left[J_{\mathrm{r}}\theta(t-\tau_{\mathrm{abs}})\exp(-(t-\tau_{\mathrm{abs}})/\tau_{\mathrm{m}})+\theta(\tau_{\mathrm{abs}}-t)D\right]$
mainly affected by the last spike, and a long-range adaptation kernel
$\eta_{\mathrm{a}}(t)=J_{\mathrm{a}}\theta(t)\exp(-t/\tau_{\mathrm{a}})$
that accumulates the spike history on a longer time scale $\tau_{\mathrm{a}}$.
An absolute refractory period is included in $\eta_{\mathrm{r}}(t)$
by setting it to $D=10^{12}$ for $0<t<\tau_{\mathrm{abs}}$. Our
choice of the kernels corresponds to a leaky IF model with dynamic
threshold \citep{Liu2001} and a reset by a constant amount $-J_{\mathrm{r}}$
after each spike. 

The parameter $c$ in Eq.~(\ref{eq:rho_esc_noise_full}) sets a baseline
firing rate and $\delta u$ sets the strength of intrinsic noise (``softness''
of threshold).  Fits of this model to pyramidal neuron recordings
yielded $\delta u\approx4\mathrm{mV}$ \citep{Jolivet2006}. Our standard
parameter set given below corresponds to an amplitude of single post-synaptic
potentials of $0.25\mathrm{mV}$ (excitatory, exc.) and $-1.1\mathrm{mV}$
(inhibitory, inh.). In the following, we measure voltage in units
of $\delta u$, so that $\delta u=1$ in dimensionless units. For
the synaptic weights $w_{ij}$, we use a homogeneous random network
as specified in Appendix~\ref{sec:Network-simulations}, below.  

The dependence of the term $\exp(-\vartheta_{i})$ in Eq.~(\ref{eq:rho_esc_noise_full})
which describes the feedback of the neuron's own spiking history $s_{i}$
can be approximated by the explicit contribution of the last spike
of the neuron at $\hat{t}$ and the average effect of previous spikes
up to $\hat{t}$ \citep{Naud2012}: 
\begin{eqnarray*}
e^{-\vartheta_{i}(t)} & \approx & e^{-\eta(t-\hat{t})}\langle e^{-\int_{-\infty}^{\hat{t}}s_{i}(t')\eta(t-t')\,\mathrm{d}t'}\rangle_{t_{i}^{k}<\hat{t}}.
\end{eqnarray*}
Here, the average is taken over all previous spike times $t_{i}^{k}<\hat{t}$.
As shown in \citep{Naud2012}, this average can be approximated by
$\exp\left[\int_{-\infty}^{\hat{t}}\left(e^{-\eta(t-t')}-1\right)\langle s_{i}(t')\rangle\, dt'\right]$.
Replacing further the firing rate $\langle s_{i}(t')\rangle$ by the
population activity $A(t')$, the threshold $\vartheta_{i}(t)$ becomes
\begin{align}
\vartheta(t,\hat{t}) & =\eta(t-\hat{t})+(\gamma_{t-\hat{t}}\ast A)(t),\label{eq:def_theta_QR}
\end{align}
for all neurons with last spike at $\hat{t}$. The kernel $\gamma_{\tau}(s)=\theta(s-\tau)(1-e^{-\eta(s)})$
represents the effect of adaptation in the quasi-renewal (QR) approximation.
Furthermore, for homogeneous random networks and large $N$, the local
field $h_{i}$ caused by synaptic input to neuron $i$ is determined
by $A$ and an effective weight $\bar{w}=N^{-2}\sum_{i,j}w_{ij}$
\citep{Brunel1999}, hence
\begin{align}
h(t) & =[\kappa\ast(J_{\mathrm{s}}A+I)](t),\label{eq:pseudo_pot}
\end{align}
where $J_{\mathrm{s}}=N\bar{w}$. The above steps enable us to treat
neuronal adaptation and network coupling in a quasi-renewal framework
with hazard function
\begin{align}
\rho_{\mathcal{H}}(t,\hat{t}) & =c\exp\left[h(t)-\vartheta(t,\hat{t})\right].\label{eq:rho_esc_noise_QR}
\end{align}
Note that $\rho_{\mathcal{H}}(t,\hat{t})$ is identical for all neurons
which have fired their last spike at $\hat{t}$. 

\begin{figure}[t]
\begin{centering}
\includegraphics[width=1\columnwidth]{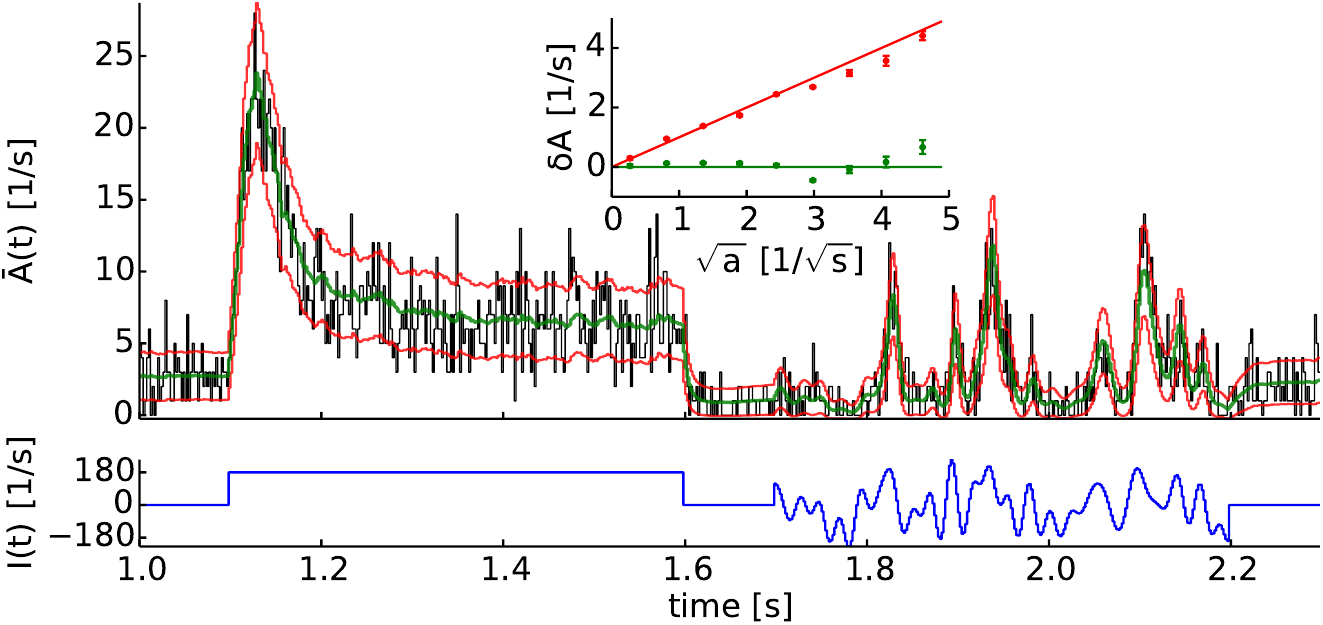}
\par\end{centering}

\caption{(Color) Quasi-renewal approximation of the population dynamics. Coarse-grained
population activity $\bar{A}(t)$ ($\bar{A}(t)=n_{0}(t)/(N\Delta t)$,
where $n_{0}(t)$ is the spike count in $[t,t+\Delta t)$, with $\Delta t=2\mathrm{ms}$,
black) resulting from 500 randomly connected adapting neurons (\ref{eq:rho_esc_noise_full})
receiving a common input current ($I(t)$, blue). The theoretical
expectation $a(t)$ (green, Eq.~\ref{eq:pop_int_abar}) based on
QR approximation (\ref{eq:rho_esc_noise_QR}), and expected fluctuations
(red, one std., $a(t)\pm(N\Delta t)^{-\frac{1}{2}}a(t)^{\frac{1}{2}}$).
At time $t$, $a(t)$ depends on the actual history $\bar{A}(t^{\prime})$
(black) for $t^{\prime}<t$. Standard parameters (see Appendix~\ref{sec:Network-simulations}),
except for $I(t)$ as shown and $c=5\mathrm{s}^{-1}$. Inset shows
mean (green) and std. (red) of deviations $\delta A(t)=\bar{A}(t)-a(t)$
as a function of $\sqrt{a}$, averaged over $25$ repetitions of the
displayed $I(t)$ (dots) vs. theory (lines).\label{fig:sim_firing_rate}}
\end{figure}
In Fig.~\ref{fig:sim_firing_rate} the population activity of a spiking
neural network simulation is compared to the theoretical prediction
(\ref{eq:pop_dyn}). To evaluate (\ref{eq:pop_int_abar}) numerically,
we iteratively compute $S_{\mathcal{H}}(t+\Delta t,\hat{t})=S_{\mathcal{H}}(t,\hat{t})(1-\rho_{\mathcal{H}}(t,\hat{t})\Delta t)$,
with $S_{\mathcal{H}}(t,t)=1$, and use $P_{\mathcal{H}}=\rho_{\mathcal{H}}S_{\mathcal{H}}$.
The QR population integral describes the response of the population
activity and its fluctuations for stationary, as well as for slowly
or rapidly varying inputs.

\subsection{Linearized population dynamics.~}

The amount of information transmitted and processed in sensory areas
of the brain is limited by the fluctuations of the population activities
\citep{Zohary1994,Mar1999,Sompolinsky2001}. Likewise, in decision
networks, finite-size induced fluctuations determine the reliability
of decisions \citep{Wang2002a,Deco2006}. The spontaneous activity
of cortical networks is typically asynchronous, and is believed to
underlie cortical information processing \citep{Renart2010}. In order
to analytically determine the power spectrum of the spontaneous population
activity, we linearize the dynamics around the large $N$ limit. To
this end, we assume that in the limit $N\rightarrow\infty$ and for
constant external input $I(t)=I_{0}$ the network dynamics has an
equilibrium point with activity $A_{0}$ corresponding to an asynchronous
firing state. With this equilibrium activity we can associate a renewal
neuron model that is obtained from the original model, Eq.~(\ref{eq:rho_esc_noise_full}),
by replacing $h_{i}(t)$ and $\vartheta_{i}(t,\hat{t})$ by $h_{0}=\kappa\ast(JA_{0}+I_{0})$
and $\vartheta_{0}(t-\hat{t})=\eta(t-\hat{t})+(\gamma_{t-\hat{t}}\ast A_{0})$,
respectively. In the following, we will use the subscript ``$0$''
to refer to quantities of the associated renewal model. For finite
system size, $N<\infty$, the activity will deviate from $A_{0}$.
Through (\ref{eq:def_theta_QR}) and (\ref{eq:pseudo_pot}) the fluctuations
$\Delta A(t)=A(t)-A_{0}$ also lead to fluctuations $\Delta h(t)=h(t)-h_{0}(t)$
and $\Delta\vartheta(t,\hat{t})=\vartheta(t,\hat{t})-\vartheta_{0}(t-\hat{t})$
, which in turn influence $A(t)$. Our goal is to determine the
spectral properties of $\Delta A(t)$. 

To simplify the derivations, we approximate the QR kernel $\gamma_{\tau}$
by its average over the inter-spike-interval density $P_{0}(\tau)$
\footnote{An alternative would be to average $\gamma_{\tau}(s)$ over the backward
recurrence time $A_{0}S_{0}(\tau)$.%
}, 
\begin{eqnarray}
\gamma(s) & = & \int_{0}^{\infty}P_{0}(\tau)\gamma_{\tau}(s)\mathrm{d}\tau=(1-e^{-\eta(s)})(1-S_{0}(s))\,.\label{eq:QR_kernel_average}
\end{eqnarray}
Expanding (\ref{eq:pop_int_abar})-(\ref{eq:corr_xi}) to first
order in $\Delta A$, $\Delta h$ and $\Delta\vartheta$, yields the
linearized stochastic dynamics (see Appendix~\ref{sec:derivation-lin})
\begin{subequations}\label{eq:pop_dyn_stat}
\begin{eqnarray}
A(t) & = & A_{0}+(Q\ast\Delta A)(t)+\sqrt{A_{0}/N}\xi(t).\label{eq:pop_dyn_A_stat}
\end{eqnarray}
Here, $Q(t)=P_{0}(t)+A_{0}\frac{\mathrm{d}}{\mathrm{d}t}(\mathcal{L}\ast[\kappa J_{\mathrm{s}}-\gamma])(t)$
determines the linear response of the expected activity $a$ to a
perturbation $\Delta A$. For our model (\ref{eq:rho_esc_noise_full}),
the kernel $\mathcal{L}$ in this expression is given by $\mathcal{L}(t)=\theta(t)\int_{0}^{\infty}\rho_{0}(s)S_{0}(s+t)\mathrm{d}s$
but $\mathcal{L}$ can be derived for most common neuron models \citep{Gerstner2000},
or, alternatively, may be estimated from neural recordings. The noise
term $\xi(t)$ is stationary Gaussian noise with correlation function
\begin{align}
\langle\xi(t)\xi(t+\tau)\rangle & =\delta(\tau)-\int_{0}^{\infty}P_{0}(s+\tau)P_{0}(s)\mathrm{d}s\label{eq:corr_xi_stat}
\end{align}
\end{subequations}for all $\tau$; cf. (\ref{eq:corr_xi}).  Eq.~(\ref{eq:pop_dyn_stat})
shows that the population activity in the stationary state is a Gaussian
process with memory, where finite-size fluctuations are described
by the colored noise $\xi(t)$.

\subsection{Fluctuations in coupled populations.~}

\begin{figure*}
\begin{centering}
\includegraphics[width=0.8\textwidth]{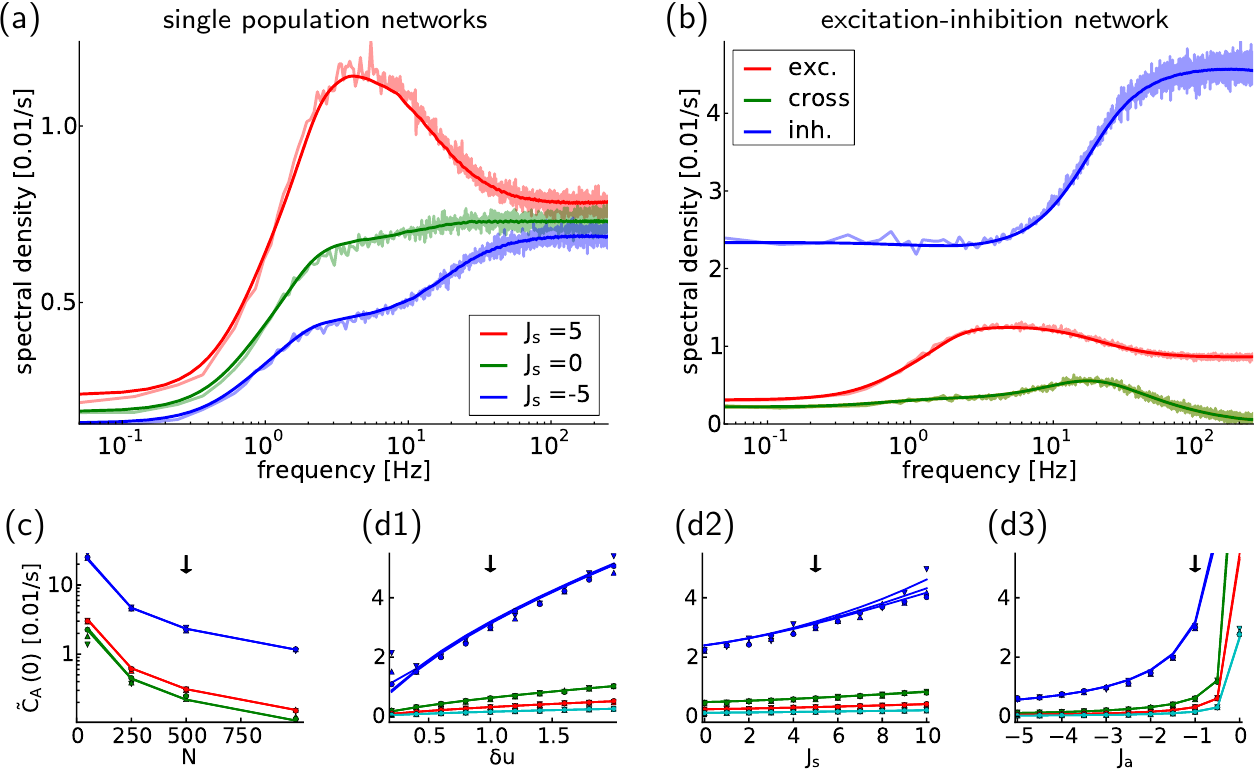}
\par\end{centering}

\centering{}\caption{(Color) Spectral density of the population activity in random networks.
Simulation (light colored lines) vs. theory (\ref{eq:spectral_density_trou})
(solid lines). (a): single populations ($K=1$), (b-d): coupled exc.-inh.
network ($K=2$), (in (b) green lines show the amplitude of the cross-spectrum).
(c-d): The limit $\omega\to0$ in dependence on parameters for the
$K=2$ case shown in (b); symbols mark simulation results for different
connection probabilities ($\bigtriangledown,\bigtriangleup,\circ$:
$p=0.2,\,0.5,\,1$) while keeping $J_{\mathrm{s}}=N\bar{w}$ fixed,
solid lines show theory (\ref{eq:spectral_density_trou}). Symbols
mostly fall on top of each other indicating independence of results
with respect to $p$. In (c) colors are as in (b), in (d1-d3) only
exc. population is shown, colors denote number of neurons (blue, green,
red, cyan: $N=50,\,250,\,500,\,1000)$. Standard parameters (marked
by arrows in (c-d), cf. Appendix~\ref{sec:Network-simulations})
were used except as indicated. \label{fig:sim_spectra} }
\end{figure*}

Let us now turn to $K$ populations consisting of $\vec{N}=(N_{1},\dots,N_{K})$
neurons. Parameters of neurons and coupling are homogeneous within
each population but may differ between one group and the next. To
incorporate network coupling, the network input (\ref{eq:pseudo_pot})
becomes $h{}_{k}(t)=(\kappa_{k}\ast(\mathbf{J}\vec{A})_{k})(t)$ for
$k=1,\dotsc,K$, where $\mathbf{J}$ is the coupling matrix and $\vec{A}(t)$
is a vector of population activities. For each population the dynamics
are given by (\ref{eq:pop_dyn_stat}) but $\mathbf{Q}(t)$ is now
a $K\times K$ matrix of coupling kernels.  Using the Fourier transform
$\tilde{f}(\omega)=\int_{-\infty}^{\infty}f(t)e^{-i\omega t}\mathrm{d}t$,
this matrix can be written as $\tilde{\mathbf{Q}}(\omega)=\tilde{\mathbf{P}}_{0}+i\omega\mathbf{A}_{0}\tilde{\mathbf{L}}(\tilde{\mathbf{K}}\mathbf{J}-\tilde{\mathbf{G}})$,
where the matrices $\mathbf{P}_{0},\mathbf{A}_{0},\mathbf{L},\mathbf{K},\mathbf{G}$
are defined as the diagonal matrices of the vectors $\vec{P}_{0},\vec{A}_{0},\vec{\mathcal{L}},\vec{\kappa},\vec{\gamma}$,
respectively. The power spectrum, defined as the Fourier transform
of the correlation function $\mathbf{C}_{A}(\tau)=\langle\Delta\vec{A}(t)\Delta\vec{A}^{T}(t+\tau)\rangle$,
can be obtained from the transformed Eq.~(\ref{eq:pop_dyn_stat})
 as $\tilde{\mathbf{C}}_{A}(\omega)=(\mathbf{1}-\tilde{\mathbf{Q}})^{-1}\mathbf{N}^{-1}\mathbf{A}_{0}(\mathbf{1}-\tilde{\mathbf{P}}_{0}\tilde{\mathbf{P}}_{0}^{\dagger})(\mathbf{1}-\tilde{\mathbf{Q}}^{\dagger})^{-1}$.
Here, $^{\dagger}$ denotes the adjoint matrix (conjugate transpose).
 It is instructive to rewrite this expression in terms of the power
spectrum of the associated renewal model \citep{Stratonovich1963}
$\tilde{\mathbf{C}}_{0}(\omega)=\mathbf{A}_{0}(\mathbf{1}-\tilde{\mathbf{P}}_{0})^{-1}(\mathbf{1}-\tilde{\mathbf{P}}_{0}\tilde{\mathbf{P}}_{0}^{\dagger})(\mathbf{1}-\tilde{\mathbf{P}}_{0}^{\dagger})^{-1}$
as follows:
\begin{align}
\tilde{\mathbf{C}}_{A}(\omega) & =\tilde{\mathbf{B}}\mathbf{N}^{-1}\tilde{\mathbf{C}}_{0}\tilde{\mathbf{B}}^{\dagger},\,\,\,\,\,\,\tilde{\mathbf{B}}=\left[\mathbf{1}-\tilde{\mathbf{R}}_{0}(\mathbf{J}-\tilde{\mathbf{K}}^{-1}\tilde{\mathbf{G}})\right]^{-1}\label{eq:spectral_density_trou}
\end{align}
Here, $\tilde{\mathbf{R}}_{0}=i\omega(\mathbf{1}-\tilde{\mathbf{P}}_{0})^{-1}\tilde{\mathbf{A}}_{0}\tilde{\mathbf{L}}\tilde{\mathbf{K}}$
is the diagonal matrix containing the linear response functions of
the associated renewal models with respect to current perturbations
$\vec{I}(t)$ \citep{Gerstner2000}. Eq.~(\ref{eq:spectral_density_trou})
shows that finite-size fluctuations are characterized by the renewal
spectrum $\tilde{\mathbf{C}}_{0}(\omega)$, shaped by recurrent input
(via $\mathbf{J}$ and $\tilde{\mathbf{K}}$) and adaptation (via
$\tilde{\mathbf{G}}$), and reduced by the factor $\mathbf{N}^{-1}$.
There are several known limit cases: first, for vanishing adaptation,
$\tilde{\mathbf{G}}=0$, we recover the linear response result of
\citep{Lindner2005,Akerberg2009,Trousdale2012} for networks of white-noise
driven IF neurons.  Our formula shows that adaptation appears as
an additional diagonal term in the effective coupling matrix $\mathbf{J}-\tilde{\mathbf{K}}^{-1}\tilde{\mathbf{G}}$,
and hence can be interpreted as an inhibitory self-coupling. Second,
if both adaptation and recurrent connections vanish, $\tilde{\mathbf{G}}=0$
and $\mathbf{J}=0$, we arrive at $\tilde{\mathbf{C}}_{A}(\omega)=\mathbf{N}^{-1}\tilde{\mathbf{C}}_{0}(\omega)$
because the superposition of independent spike trains does not change
the shape of the power spectrum. Third, our result also includes
the frequently employed Hawkes process \citep{Hawkes1971a,Pernice2012,Helias2013},
which is recovered for a constant single neuron spectrum $\tilde{\mathbf{C}}_{0}=\mathbf{A}_{0}$
and vanishing adaptation, $\tilde{\mathbf{G}}=0$.  For a comparison
of our result to simulations, see Fig.~\ref{fig:sim_spectra}, which
will be discussed below. In section \ref{sub:info} we make use of
Eq.~(\ref{eq:spectral_density_trou}) to quantify information filtering
in neural populations (Fig.~\ref{fig:sim_info}). A comparison to
the special cases of Eq.~(\ref{eq:spectral_density_trou}) described
above is shown in Fig.~\ref{fig:sim_spectra_SI}. 

In Fig.~\ref{fig:sim_spectra}(a)-(b) the spectral density is shown
compared to simulations, where the frequency equals $\omega/(2\pi)$.
The spectra are well described by the novel theory, which captures
refractoriness, recurrent feedback and the reduction of power at
low frequencies due to adaptation. The latter arises from negative
correlations between ISIs typical for adapting neurons \citep{Schwalger2013}.
Interestingly, this purely non-renewal effect is well accounted for
by our quasi-renewal theory. Since adaptation effects are most prominent
at low frequency, we examined the dependence of the power on the model
parameters for $\omega\to0$ (Fig.~\ref{fig:sim_spectra}(c)-(d)).
Our theory describes the simulations well across the studied parameter
range.

\subsection{Influence of correlated external signals.~}

Neuronal networks in the brain are subject to external influences,
either due to sensory input or ongoing activity in other brain areas.
How do neuronal populations respond to small, time-dependent input
currents $\vec{I}(t)=(I_{1},\dots,I_{L})$ with spectral density $\tilde{\mathbf{C}}_{I}(\omega)$?
To answer this question, we proceed as before and linearize Eqs.~(\ref{eq:pop_int_abar})-(\ref{eq:corr_xi})
with respect to the small fluctuations $\Delta h_{k}(t)=(\kappa_{k}\ast(\mathbf{J}\Delta\vec{A}+\mathbf{M}\vec{I})_{k})(t)$
of the local field. Here, we restrict our analysis to independent
inputs $\vec{I}$, such that $\tilde{\mathbf{C}}_{I}$ is diagonal
with entries $\tilde{C}_{I,i}$, but also included a $K\times L$
mixing matrix $\mathbf{M}$, which allows us to model shared input.
The resulting spectral density is given by the sum 
\begin{align}
\tilde{\mathbf{C}}_{A}(\omega)= & \tilde{\mathbf{B}}[\mathbf{N}^{-1}\tilde{\mathbf{C}}_{0}+\tilde{\mathbf{R}}_{0}\mathbf{M}\tilde{\mathbf{C}}_{I}\mathbf{M}^{\dagger}\tilde{\mathbf{R}}_{0}^{\dagger}]\tilde{\mathbf{B}}^{\dagger}.\label{eq:spectral_density_iext}
\end{align}
Thus, additional fluctuations due to the stimulus $\vec{I}$ are shaped
by both the single neuron filter $\mathbf{\tilde{R}}_{0}\mathbf{M}$
and the network and adaptation filter $\mathbf{\tilde{B}}$ (\ref{eq:spectral_density_trou}),
combining the effects of recurrent connectivity and adaptation.

\subsection{Information transmission.~\label{sub:info}}

Our theory allows us to quantify the transmission of information from
external input signals $\vec{I}(t)$ through a system of coupled neural
populations. The coherence between the signal $j$ and the activity
of population $i$
\begin{equation}
\Gamma_{ij}(\omega)=\frac{|\langle\tilde{I}_{j}^{\ast}(\omega)\tilde{A}_{i}(\omega)\rangle|^{2}}{\tilde{C}_{I,j}(\omega)(\tilde{\mathbf{C}}_{A}(\omega))_{ii}}\label{eq:def_coherence}
\end{equation}
\textcolor{blue}{}can be regarded as a frequency resolved measure
of information transmission. Information theory \citep{Shannon1948,Gabbiani1996,Borst1999}
states that the mutual information rate is bounded from below by $-\int_{0}^{\infty}\log_{2}[1-\Gamma_{ij}(\omega)]\frac{\mathrm{d}\omega}{2\pi}$. 

Since adaptation attenuates the response to slowly changing signals,
one might expect that it also attenuates low frequency information
content. For a single population, however, this is not the case, but
instead the coherence is low-pass, i.e. it monotonically decreases
for increasing frequency \citep{Akerberg2009}. Here we show that
in coupled populations of adapting neurons, coherences can be non-monotonic
allowing the neural circuit to preferentially encode information in
certain frequency bands. Put differently, a multi-population setup
can realize an information filter.

Using Eq.~(\ref{eq:def_coherence}), we find the general form of
the coherence matrix: 
\begin{align}
\Gamma_{ij}(\omega) & =\frac{|(\tilde{\mathbf{B}}\tilde{\mathbf{R}}_{0}\mathbf{M})_{ij}|^{2}\tilde{C}_{I,j}}{\sum_{k=1}^{K}|\tilde{B}_{ik}|^{2}\tilde{C}_{0,k}+\sum_{l=1}^{L}|(\tilde{\mathbf{B}}\tilde{\mathbf{R}}_{0}\mathbf{M})_{il}|^{2}\tilde{C}_{I,l}}\,.\label{eq:coherence_iext}
\end{align}
In this expression, the numerator represents the contribution of
the signal $I_{j}(t)$ to the power spectrum of population $i$. This
effective signal power is divided by the total power spectrum of population
$i$, which consists of direct ($k=i$) and indirect ($k\ne i$) sources
of variability. Both sources contain internally generated noise due
to finite size $N_{k}$ as well as signal power. However, the diagonal
elements ($k=i$) of the shaping matrices can be much stronger than
the off-diagonal elements ($k\ne i$) depending on the coupling matrix
$\mathbf{J}$. Therefore, we expect that the direct source of variability
dominates in the denominator.

If there is only one population and signal ($K=1$, $L=1$), the term
$|\tilde{B}_{11}(\omega)|^{2}$ occurs in both numerator and denominator
and cancels. Thus, coupling and adaptation do not shape the coherence
in a single population. Furthermore, the signal term $|\tilde{R}_{0,1}(\omega)|^{2}M_{11}\tilde{C}_{I,1}(\omega)$
is matched in both numerator and denominator, which leads to a flat
coherence at frequencies where the signal dominates the finite $N$
noise. At high frequencies, the neural response amplitude $|\tilde{\mathbf{R}}_{0}|^{2}$
decays due to the leaky membrane, but the spontaneous spectrum $\tilde{\mathbf{C}}_{0}$
has a constant high-frequency limit equal to $\mathbf{A}_{0}$. One
therefore typically observes a low-pass like information transfer
characteristics of single neurons or populations \citep{Chacron2004,Vilela2009,Akerberg2009}. 

For several populations ($K>1$), however, we can distinguish two
cases: If the signal is read out at a receiving population, $M_{ij}\neq0$,
the signal power in the numerator is matched by the dominating direct
signal power in the denominator, and hence the shaping of the signal
power cancels. In contrast, if read out at a different population,
$M_{ij}=0$, the signal $j$ contributes only indirectly to the power
spectrum of population $i$ via synaptic connections. Thus, we expect
that the shape of signal power and power spectrum (i.e. numerator
and denominator, respectively) is generally different if the transmission
path involves multiple populations. 

\begin{figure}
\begin{centering}
\includegraphics[width=1\columnwidth]{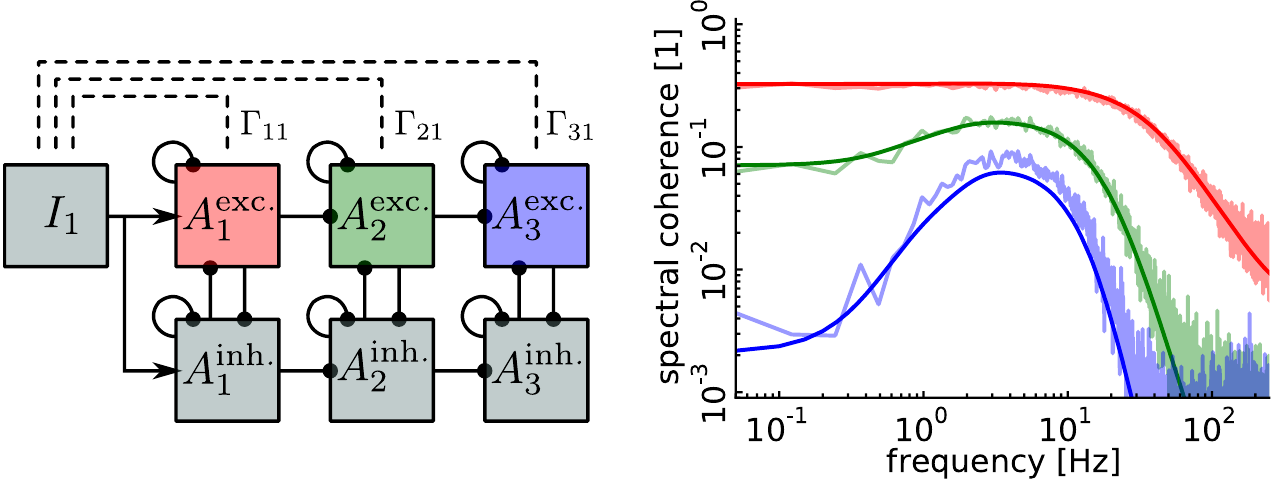}
\par\end{centering}

\caption{(Color) Signal processing and coherence shaping in a feed-forward
chain of recurrently connected neural populations, simulation vs.
theory. Left: Schematic of the network. Right: spectral coherences
$\Gamma_{i1}$ (\ref{eq:coherence_iext}) of $I_{1}$ and $A_{i}^{\mathrm{exc.}}$,
theory (\ref{eq:coherence_iext}) (lines) vs. simulation result (light
colored lines). Colors indicate coherences of signals $I_{1}$ and
population activity $A_{i}$. The mutual information rate is closely
related to the spectral coherence, see text. Standard parameters,
except for increased coupling $J_{\mathrm{s}}=10$. The input current
$I_{1}(t)$ is a Gaussian white noise with spectral density $\tilde{C}_{I,1}(\omega)=9\mathrm{s}^{-1}$.
\label{fig:sim_info}}
\end{figure}
As an example of this mechanism, we show a feed-forward chain
of excitatory and inhibitory populations (Fig.~\ref{fig:sim_info},
$K=6$, $L=1$). In the first layer, the effective signal power is
reduced at low frequencies  because of adaptation and inhibitory
feedback. However, the power spectrum of $A_{\mathrm{exc},1}(t)$
is dominated by the same signal power and hence exhibits a very similar
reduction of low-frequency power. Consequently, the coherence (being
the ratio of these two spectra) is rather flat at low frequencies
and shows a decay at higher frequencies (Fig.~\ref{fig:sim_info},
red lines). This low-pass characteristics changes at later stages
in the chain (green \& blue): The signal term  is increasingly more
shaped by adaptation and coupling properties, whereas the noise spectrum
changes less. As a result, the coherence shows a maximum at a finite
frequency (Fig.~\ref{fig:sim_info}, blue lines). This band-pass
structure becomes more pronounced from layer to layer, representing
a form of information filtering. Coherence functions with band-pass
characteristics have been observed in neurons postsynaptic to electro-receptor
afferents in electric fish \citep{Chacron2005b,Krahe2008}.

\section{Conclusions}

We have shown that fluctuations in finite-sized networks of spiking
neurons are captured by a colored noise term added to the population
integral equation of the infinite system. Our approach yields spectral
densities of the population activity in randomly or fully connected
multi-population networks which are in excellent agreement with simulation
results. Our quasi-renewal theory includes refractory effects and
adaptation on multiple time scales. In contrast to earlier treatments
of neuronal refractory effects in population dynamics \citep{Toyoizumi2009,Pernice2012,Helias2013}
or linear response formulas for adaptive neurons \citep{Richardson2009},
the QR population integral, derived directly from the neuron model
definition, captures the time-dependent, non-linear dynamics and adaptation
of neural population activity. 

We applied our theory to information filtering by coupled populations
of spiking neurons with adaptation. We showed that, although impossible
in single populations due to a cancellation of signal and noise terms
of the coherence, coupled populations can filter information through
adaptation mechanisms and neuronal interactions. This mechanism might
be exploited in the layered structure of cortical circuits, or in
sensory systems of insects where signals traverse a sequence of nuclei.

In this paper we treated populations of point neurons with static
synapses and applied linear response theory. How to generalize our
theory to incorporate effects of nonlinear dendritic integration,
spike-synchrony detection and short-term synaptic plasticity, which
all contribute to information filtering \citep{Rosenbaum2012,Sharafi2013,Droste2013},
is an important question that merits further investigation. Nonetheless,
due to the versatility of GLM models, our theory already provides
a useful tool for interpreting neural data at the population level.
For example, our theory suggests that in \textit{in-vitro} experiments
with optogenetically evoked input currents and simultaneous measurements
of neural activity \citep{ElHady2013}, system parameters may be identified
based on the relation of the spectra, Eq.~(\ref{eq:spectral_density_iext}).
Moreover, large-scale neural systems can now be analyzed as coupled
populations of model neurons with single-cell parameters extracted
from experiments, and simulated using a muti-scale approach.

\section*{Acknowledgements}

Research was supported by the European Research Council (no. 268 689,
T.~Schwalger and W.~Gerstner) and by the Swiss National Science
Foundation (no. 200020\_147200, M.~Deger). We thank Laureline Logiaco
for helpful discussions. 

M.~D. and T.~S. contributed equally to this work.

\appendix

\section{Network simulations \label{sec:Network-simulations}}

We compare our theoretical results to simulations of networks of
excitatory and inhibitory neurons defined by (\ref{eq:rho_esc_noise_full}).
In the case $K=1$ (Fig.~\ref{fig:sim_spectra}(a)), we use $\vec{N}=N$
and $\mathbf{J}=J_{\mathrm{s}}$, in the case $K=2$ (Fig.~\ref{fig:sim_spectra}(b)-(d)),
$\vec{N}=(\nicefrac{4}{5}N,\nicefrac{1}{5}N)$, and $\mathbf{J}=((J_{\mathrm{s}},-1.1\, J_{\mathrm{s}}),(J_{\mathrm{s}},-1.1\, J_{\mathrm{s}}))$.
In the case $K=6$ (Fig.~\ref{fig:sim_info}), each exc. (inh.) population
consists of $\nicefrac{4}{5}N$ (\foreignlanguage{english}{$\nicefrac{1}{5}N$)}
neurons and $\mathbf{J}$ with entries as in Fig.~\ref{fig:sim_info}A,
where exc. (inh.) couplings are $J_{\mathrm{s}}$ ($-1.1\, J_{\mathrm{s}}$).
Generally, neurons of population $i$ receive synapses from a random
subset of $pN_{j}$ neurons of population $j$, each with synaptic
weight $w_{ij}=J_{ij}/(pN_{j})$ and delay $\tau_{\mathrm{s}}$. Unless
the connection probability $p$ is $1$, self-connections are excluded.
 Networks were simulated for $2\cdot10^{4}\mathrm{s}$ using NEST
\citep{Gewaltig:NEST} (neuron model pp\_psc\_delta, temporal resolution
$2\mathrm{ms}$). Standard parameters, unless indicated otherwise:
$N=500$, $c=10\mathrm{s}^{-1}$, $\tau_{\mathrm{m}}=0.01\mathrm{s}$,
$\tau_{\mathrm{a}}=0.3\mathrm{s}$, $\Delta t=\tau_{\mathrm{s}}=\tau_{\mathrm{abs}}=2\mathrm{ms}$,
$J_{\mathrm{r}}=3$, $J_{\mathrm{a}}=1$, $J_{\mathrm{s}}=5$, $p=0.2$,
$I_{0}=0,$ $\tilde{C}_{I}=0$. In the exc.-inh. network, for the
inh. neurons which typically show little adaptation \citep{Mensi2012}
we deactivated adaptation by setting $J_{\mathrm{a}}=0$ and $c=5\mathrm{s}^{-1}$.
While it is possible to theoretically approximate the stationary interval
distribution $P_{0}$ by searching for a self-consistent rate $A_{0}$
as described in \citep{Naud2012}, here we use $P_{0}$ from simulated
inter-spike-intervals of each population. From the measured $P_{0}(t)$
we derive $A_{0}=1/\int_{0}^{\infty}tP_{0}(t)\mathrm{d}t$, $S_{0}(t)=\int_{t}^{\infty}P_{0}(t^{\prime})\mathrm{d}t^{\prime}$
and $\rho_{0}(t)=P_{0}(t)/S_{0}(t)$.
\begin{figure*}
\begin{centering}
\includegraphics[width=0.8\textwidth]{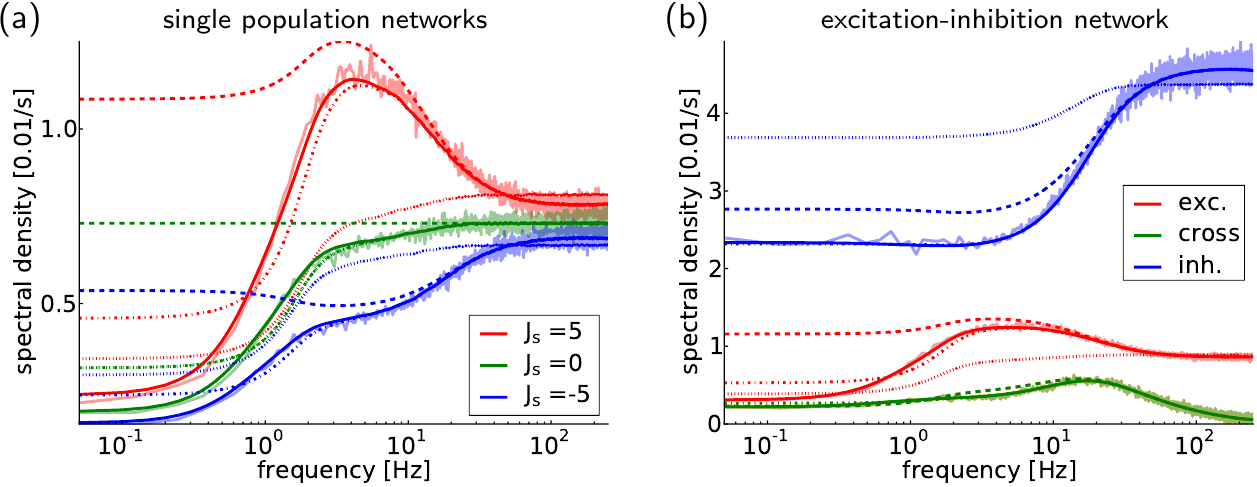}\vspace{0.1cm}

\par\end{centering}

\centering{}\caption{(Color) Spectral density of the population activity in random networks
(as Fig.~\ref{fig:sim_spectra}(a)-(b)) with comparison to earlier
theories (special cases). Simulation (light colored lines) vs. theory
(\ref{eq:spectral_density_trou}) (solid lines). For comparison: uncoupled
renewal processes ($\tilde{\mathbf{B}}=\mathbf{1}$), dotted; coupled
renewal processes ($\tilde{\mathbf{B}}^{-1}=\mathbf{1}-\tilde{\mathbf{R}}_{0}\mathbf{J}$),
dash-dotted; Hawkes process ($\tilde{\mathbf{B}}^{-1}=\mathbf{1}-\tilde{\mathbf{R}}_{0}\mathbf{J}$,
$\tilde{\mathbf{C}}_{0}=\mathbf{A}_{0}$), dashed lines. (a): single
populations ($K=1$), (b): coupled exc.-inh. network ($K=2$). Blue
dash-dotted and solid lines coincide because inhibitory neurons here
have no adaptation ($J_{\mathrm{a}}=0$). Standard parameters were
used except as indicated. \label{fig:sim_spectra_SI} }
\end{figure*}

\section{Detailed derivation of Eq.~(\ref{eq:corr_xi})\label{sec:derivation-fn}}

\begin{figure}
\centering{} \includegraphics[width=1\linewidth]{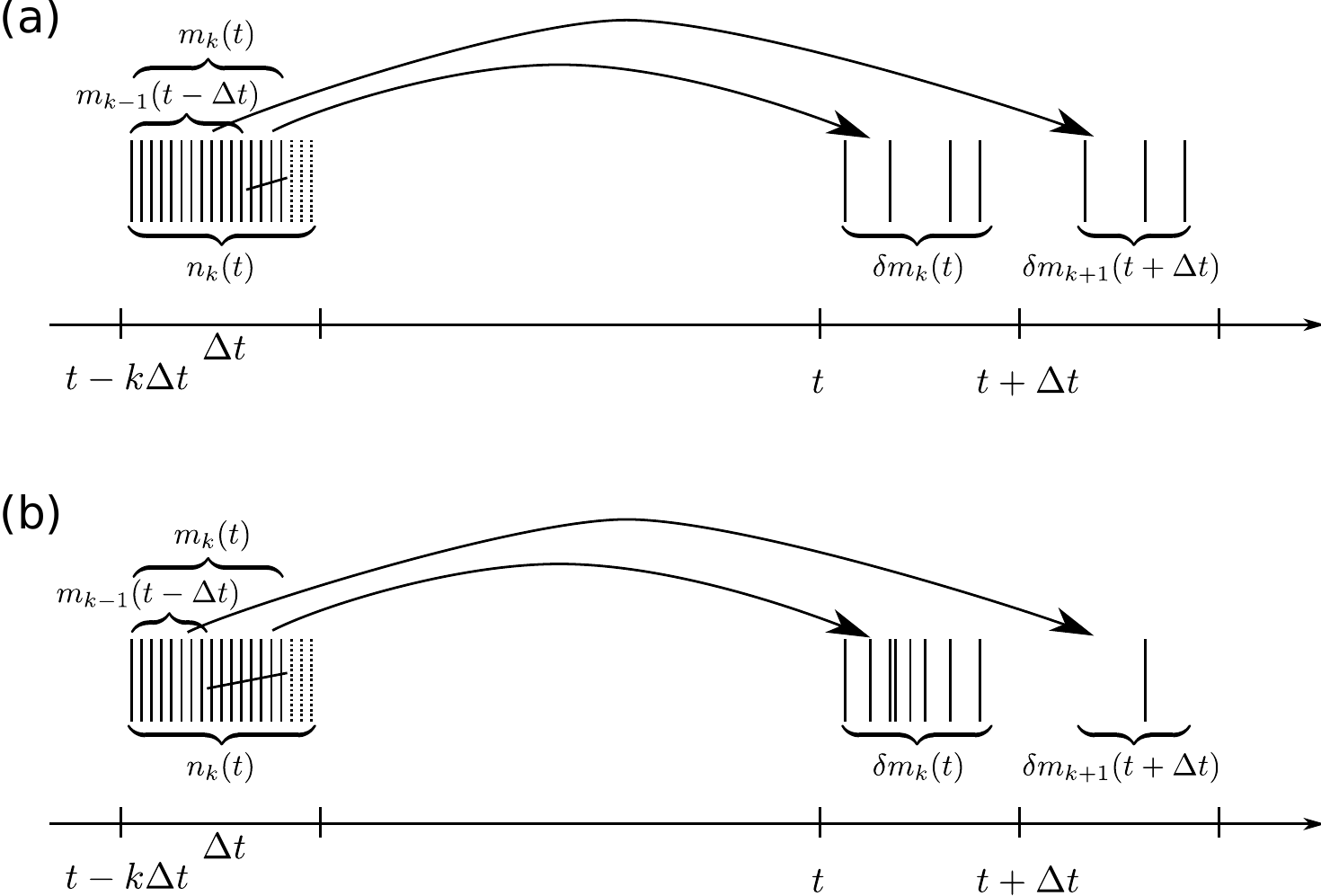}
\caption{Illustration of negative correlations between $\delta m_{k}(t)$ and
$\delta m_{k+1}(t+\Delta t)$. (a): Expected number of spikes from
group $k$ in bins $[t,t+\Delta t]$ and $[t+\Delta t,t+2\Delta t]$.
(b): A large fluctuation of $\delta m_{k}(t)$ leads to reduction
of the number of available spikes $m_{k+1}(t+\Delta t)$. As a consequence,
$\delta m_{k+1}(t+1)$ tends to be small. \label{fig:corr-illu}}
\end{figure}

The aim is to find a dynamical equation for the population activity
\begin{equation}
A(t)=\frac{1}{N}\sum_{i=1}^{N}s_{i}(t)=\lim_{\Delta t\rightarrow0}\frac{n_{0}(t)}{N\Delta t},\label{eq:A}
\end{equation}
where $n_{0}(t)$ is the total number of spikes in the interval $[t,t+\Delta t]$.
More generally, we define $n_{k}(t)$, $k\in\mathbb{Z}$, as the total
number of spikes in $[t-k\Delta t,t-(k-1)\Delta t]$, i.e. the activity
$k$ time bins in the past. It is useful to consider furthermore the
total number of neurons that spiked in the time bin $[t-k\Delta t,t-(k-1)\Delta t]$,
$k=1,2,\dotsc$, but had no further spike until time $t$. Let us
denote this number by $m_{k}(t)$ (Fig.~\ref{fig:corr-illu} and
Table~\ref{tab:symbols}). The number of neurons that spike in $[t,t+\Delta t]$
and had their last spike in the bin $[t-k\Delta t,t-(k-1)\Delta t]$
shall be denoted by $\delta m_{k}(t)$. These neurons decrease the
number $m_{k}(t)$ of neurons from group $k$ that had survived until
time $t+\Delta t$ in the next time step, i.e. 
\begin{equation}
\delta m_{k}(t)=m_{k}(t)-m_{k+1}(t+\Delta t),\qquad k=1,2,\dotsc.\label{eq:xi-equa}
\end{equation}
 The total number of spikes at time $t$, $n_{0}(t)$, is the sum
over all possible last spike times, hence 
\begin{equation}
n_{0}(t)=\sum_{k=1}^{\infty}\delta m_{k}(t).\label{eq:sum}
\end{equation}

We will now express the activity $n_{0}(t)$ in terms of the past
activity $\mathcal{H}_{t}=\{n_{k}(t)\}_{k=1,2,\dotsc}$, using the
Gaussian approximation. This requires to compute the mean and correlation
function of $n_{0}(t)$ \emph{given} the past values $n_{k}(t)$,
$k=1,2,\dotsc$. In the following, the averaging bracket $\left\langle \cdot\right\rangle $
has to be understood as the conditional average $\left\langle \cdot\right\rangle _{\mathcal{H}_{t}}=\left\langle \cdot\right\rangle _{\{n_{k}(t)\}_{k=1,2,\dotsc}}$,
i.e. we will omit the conditioning subscript for simplicity. Although
$n_{k}(t)$, the total number of spikes in bin $t-k\Delta t$, is
fixed, the number $m_{k}(t)$ of neurons that had their \emph{last}
spike in bin $t-k\Delta t$ is variable. It is this variability that
we will average over (This corresponds to a statistical ensemble of
populations that all have an identical history of population activity
$n_{k}(t)$, $k=1,2,\dotsc$.). 
\begin{table}
\begin{tabular}{l|l}
$n_{0}(t)$ & \# of neurons which spike in $[t,t+\Delta t]$\tabularnewline
\hline 
$n_{k}(t)$ & \# of neurons which spiked in \tabularnewline
 & $[t-k\Delta t,t-(k-1)\Delta t]$ ; $n_{k}(t)=n_{0}(t-k\Delta t)$\tabularnewline
\hline 
$m_{k}(t)$ & \# of neurons with last spike in\tabularnewline
 & $[t-k\Delta t,t-(k-1)\Delta t]$\tabularnewline
\hline 
$\delta m_{k}(t)$ & \# of neurons with last spike in \tabularnewline
 & $[t-k\Delta t,t-(k-1)\Delta t]$ and next spike in $[t,t+\Delta t]$\tabularnewline
\hline 
$\rho(t,\hat{t})$ & hazard function: rate at $t$ given last spike at $\hat{t}$\tabularnewline
\hline 
$S(t,\hat{t})$ & survivor function: probability of no spike in $[\hat{t},t]$\tabularnewline
\hline 
$P(t,\hat{t})$ & inter-spike-interval density: probability density \tabularnewline
 & of next spike at $t$ given last spike at $\hat{t}$; ~~$P=\rho\cdot S$\tabularnewline
\end{tabular}

\caption{Definitions of symbols used in Appendix~\ref{sec:derivation-fn}.\label{tab:symbols}}
\end{table}

Suppose we know the value $m_{k}(t)$ of the group of neurons with
their last spike in $[t-k\Delta t,t-(k-1)\Delta t]$. Then the expected
number of spikes from that group in the next interval is 
\begin{align}
\langle\delta m_{k}(t)\rangle_{m_{k}(t)} & =\rho(t,t-k\Delta t)\cdot\Delta t\cdot m_{k}(t)\,,\label{eq:dmk_cond_avg}
\end{align}
where $\rho(t,\hat{t})$ is the hazard function of the neurons (instantaneous
rate at time $t$ given last spike at $\hat{t}$), and $\langle x\rangle_{y}$
denotes the expectation of $x$ conditioned on $y$ (in addition to
the overall condition of a fixed history $\mathcal{H}_{t}=\{n_{k}(t)\}_{k=1,2,\dotsc}$).
But since we do not know the exact value of $m_{k}(t)$ we need to
average 
\begin{align}
\langle\delta m_{k}(t)\rangle & =\langle\langle\delta m_{k}(t)\rangle_{m_{k}(t)}\rangle\nonumber \\
 & =\rho(t,t-k\Delta t)\cdot\Delta t\cdot\langle m_{k}(t)\rangle\label{eq:dmk_avg}
\end{align}
where we have used (\ref{eq:dmk_cond_avg}). We now use this result
to calculate the expected number of spikes in the interval $[t,t+\Delta t]$.
Averaging over (\ref{eq:sum}) yields 
\begin{align}
\langle n_{0}(t)\rangle & =\sum_{k=1}^{\infty}\rho(t,t-k\Delta t)\cdot\Delta t\cdot\langle m_{k}(t)\rangle\,.\label{eq:mean_proto}
\end{align}
The \emph{average} number of neurons that fired their last spike in
$[t-k\Delta t,t-(k-1)\Delta t]$ and survived up to $t$ can be expressed
using the survival probability $S(t,\hat{t})=\exp(-\int_{\hat{t}}^{t}\rho(t^{\prime},\hat{t})dt^{\prime})$
as follows: 
\begin{align}
\langle m_{k}(t)\rangle & =S(t,t-k\Delta t)n_{k}(t)\nonumber \\
 & =S(t,t-k\Delta t)n_{0}(t-k\Delta t)\,.\label{eq:mk_avg_surv}
\end{align}
We can now take the limit $\Delta t\to0$ in (\ref{eq:mean_proto})
and find 
\begin{align}
\frac{\langle n_{0}(t)\rangle}{\Delta t} & \to N\int_{-\infty}^{t}P(t,\hat{t})A(\hat{t})\,\mathrm{d}t',\quad\Delta t\to0,\label{eq:mean}
\end{align}
where $P(t,\hat{t})=\rho(t,\hat{t})S(t,\hat{t})$ is the inter-spike-interval
density. Eq.~(\ref{eq:mean}) is equivalent to Eq.~(\ref{eq:pop_int_abar}).

To obtain the correlation function we can write for $q\in\mathbb{Z}$
\begin{equation}
\langle n_{0}(t)n_{0}(t+q\Delta t)\rangle=\sum_{k,l=1}^{\infty}\langle\delta m_{k}(t)\delta m_{l}(t+q\Delta t)\rangle.\label{eq:corr-func}
\end{equation}
Here, the spike numbers $\delta m_{k}(t)$ and $\delta m_{l}(t+q\Delta t)$
that refer to different groups $k$ and $l-q$ are uncorrelated. Correlations
only arise for $\delta m_{k}(t)$ and $\delta m_{k+q}(t+q\Delta t)$,
i.e. spikes that refer to the same group in the past. Thus, 
\begin{multline}
\langle n_{0}(t)n_{0}(t+q\Delta t)\rangle=\sum_{k,l=1}^{\infty}\langle\delta m_{k}(t)\rangle\langle\delta m_{l}(t+q\Delta t)\rangle\\
+\sum_{k=1}^{\infty}\langle\delta m_{k}(t)\delta m_{k+q}(t+q\Delta t)\rangle\\
-\sum_{k=1}^{\infty}\langle\delta m_{k}(t)\rangle\langle\delta m_{k+q}(t+q\Delta t)\rangle,
\end{multline}
so that the covariance is 
\begin{multline}
\langle\Delta n_{0}(t)\Delta n_{0}(t+q\Delta t)\rangle=\sum_{k=1}^{\infty}\langle\delta m_{k}(t)\delta m_{k+q}(t+q\Delta t)\rangle\\
-\sum_{k=1}^{\infty}\langle\delta m_{k}(t)\rangle\langle\delta m_{k+q}(t+q\Delta t)\rangle,\label{eq:cov}
\end{multline}
where $\Delta n_{0}(t)=n_{0}(t)-\langle n_{0}(t)\rangle$. Therefore
we need to compute 
\begin{align}
\langle\delta m_{k}(t)\delta m_{k+q}(t+q\Delta t)\rangle\,.\label{eq:corr_expression}
\end{align}
To this end, let us consider the cases $q=0$ and $q>0$ separately. 

For $q=0$ and large $N$, the number of neurons that spike in $[t,t+\Delta t]$
and had their last spike at $t-k\Delta t$ is a Poisson variable with
mean and variance $\Delta t\, P(t,\, t-k\Delta t)\, n_{k}(t)$. Thus
(\ref{eq:corr_expression}) becomes 
\begin{equation}
\left\langle [\delta m_{k}(t)]^{2}\right\rangle =\Delta t\, P(t,\, t-k\Delta t)\, n_{k}(t)+\mathcal{O}(\Delta t^{3}).\label{eq:xi-var}
\end{equation}

For $q>0$, we employ (\ref{eq:xi-equa}) twice and obtain
\begin{multline}
\langle\delta m_{k}(t)\delta m_{k+q}(t+q\Delta t)\rangle=\langle m_{k}(t)\, m_{k+q}(t+q\Delta t)\rangle\\
-\langle m_{k+1}(t+\Delta t)\, m_{k+q}(t+q\Delta t)\rangle\\
-\langle m_{k}(t)\, m_{k+q+1}(t+(q+1)\Delta t)\rangle\\
+\langle m_{k+1}(t+\Delta t)\, m_{k+q+1}(t+(q+1)\Delta t)\rangle.\label{eq:xi_corr_expanded}
\end{multline}
In order to evaluate each of these four correlators, we note that
the probability that a neuron from group $k$ ``survives'' until
time $t+q\Delta t$ given that it survived until time $t$ is $S(t+q\Delta t,t-k\Delta t)/S(t,t-k\Delta t)$
according to Bayes law. Thus, out of the $m_{k}(t)$ neurons that
survived until time $t,$ on average 
\begin{multline}
\left\langle m_{k+q}(t+q\Delta t)\right\rangle {}_{m_{k}(t)}=\frac{S(t+q\Delta t,t-k\Delta t)}{S(t,t-k\Delta t)}\cdot m_{k}(t),\label{eq:mk_evol_survivor}
\end{multline}
also survive until $t+q\Delta t$. Therefore, the correlator for $0\le l<q$
can be written as
\begin{multline*}
\left\langle m_{k+l}(t+l\Delta q)m_{k+q}(t+q\Delta t)\right\rangle =\\
=\left\langle m_{k+l}(t+l\Delta q)\left\langle m_{k+q}(t+q\Delta t)\right\rangle _{m_{k+l}(t+l\Delta t)}\right\rangle ,\\
=\frac{S(t+q\Delta t,t-k\Delta t)}{S(t+l\Delta t,t-k\Delta t)}\cdot\left\langle m_{k+l}^{2}(t+l\Delta t)\right\rangle .
\end{multline*}
Applying this result to (\ref{eq:xi_corr_expanded}), we obtain
\begin{multline}
\langle\delta m_{k}(t)\delta m_{k+q}(t+q\Delta t)\rangle=\\
\left[S(t+(q+1)\Delta t,t-k\Delta t)-S(t+q\Delta t,t-k\Delta t)\right]\\
\times\left(\frac{\langle[m_{k+1}(t+\Delta t)]^{2}\rangle}{S(t+\Delta t,t-k\Delta t)}-\frac{\langle[m_{k}(t)]^{2}\rangle}{S(t,t-k\Delta t)}\right)\label{eq:xi_corr_proto}
\end{multline}

How can we calculate the second moment of $m_{k}(t)$? Recall that
$m_{k}(t)$ is the part of the $n_{k}(t)$ neurons firing in bin $t-k\Delta t$
that survived until time $t$. Thus $m_{k}(t)$ can be regarded as
a binomially distributed random number with $n=n_{k}(t)$ trials and
survival probability $p=S(t,\, t-k\Delta t)$. This random number
has mean $np$ and variance $np(1-p)$. Hence, the second moment reads
\begin{subequations} 
\begin{multline}
\langle[m_{k}(t)]^{2}\rangle=\langle[m_{k}(t)-\langle m_{k}(t)\rangle]^{2}\rangle+\underbrace{\langle m_{k}(t)\rangle^{2}}_{\mathcal{O}(\Delta t^{2})}\\
=n_{k}(t)S(t,\, t-k\Delta t)\left[1-S(t,\, t-k\Delta t)\right]+\mathcal{O}(\Delta t^{2}).\label{eq:m_k-var}
\end{multline}
Likewise,\label{m-var} 
\begin{multline}
\langle[m_{k+1}(t+\Delta t)]^{2}\rangle=n_{k}(t)S(t+\Delta t,t-k\Delta t)\\
\times\left[1-S(t+\Delta t,t-k\Delta t)\right]+\mathcal{O}(\Delta t^{2})\label{eq:m_kp1-var}
\end{multline}
\end{subequations}because $n_{k+1}(t+\Delta t)=n_{k}(t)$. Inserting
(\ref{m-var}) into (\ref{eq:xi_corr_proto}) we find 
\begin{multline}
\langle\delta m_{k}(t)\delta m_{k+q}(t+q\Delta t)\rangle=n_{k}(t)\,\Delta t^{2}\\
\times\frac{S(t+(q+1)\Delta t,t-k\Delta t)-S(t+q\Delta t,t-k\Delta t)}{\Delta t}\\
\times\frac{S(t,t-k\Delta t)-S(t+\Delta t,t-k\Delta t)}{\Delta t}\\
=-P(t+q\Delta t,t-k\Delta t)P(t,\, t-k\Delta t)\, n_{k}(t)\,\Delta t^{2}\,,.\label{eq:xi-corr2}
\end{multline}
Here, we have identified the derivative $\nicefrac{\mathrm{d}}{\mathrm{d}t}S(t,\hat{t})=-P(t,\hat{t})$.
Note that this expression is of order $\mathcal{O}(\Delta t^{3})$,
whereas $\langle\delta m_{k}(t)\rangle\langle\delta m_{k+q}(t+q\Delta t)\rangle$
is of order $\mathcal{O}(\Delta t^{4})$. So we can neglect the second
term on the right-hand side of Eq.~(\ref{eq:cov}). 

Putting all together, we find 
\begin{multline}
\frac{1}{\Delta t^{2}}\langle\Delta n_{0}(t)\Delta n_{0}(t^{\prime}=t+q\Delta t)\rangle\\
=\frac{1}{\Delta t^{2}}\sum_{k=1}^{\infty}\langle\delta m_{k}(t)\delta m_{k+q}(t+q\Delta t)\rangle\\
=\frac{1}{\Delta t}\delta_{q,0}\left[\sum_{k=1}^{\infty}P(t,\, t-k\Delta t)n_{k}(t)+\mathcal{O}(\Delta t)\right]\\
-\sum_{k=1}^{\infty}P(t+q\Delta t,\, t-k\Delta t)P(t,\, t-k\Delta t)n_{k}(t)\\
\xrightarrow[\Delta t\to0]{}\, N\delta(t-t^{\prime})\int_{-\infty}^{t}P(t,t^{\prime\prime})A(t^{\prime\prime})\,\mathrm{d}t^{\prime\prime}\\
-N\int_{-\infty}^{t}P(t,t^{\prime\prime})P(t^{\prime},t^{\prime\prime})A(t^{\prime\prime})\,\mathrm{d}t^{\prime\prime}.\label{eq:noise-cov-discrete}
\end{multline}
 Thus, using (\ref{eq:mean}), we arrive at the final result 
\begin{equation}
A(t)=\int_{-\infty}^{t}P(t,t^{\prime})A(t^{\prime})\,\mathrm{d}t^{\prime}+\delta A(t)\label{eq:final}
\end{equation}
 where $\delta A(t)$ is Gaussian with conditional correlation function
\begin{multline}
\langle\delta A(t)\delta A(t+\tau)\rangle=N^{-1}\delta(\tau)\int_{-\infty}^{\infty}P(t,t^{\prime})A(t^{\prime})\,\mathrm{d}t^{\prime}\\
-N^{-1}\int_{-\infty}^{\infty}P(t+\tau,t^{\prime})P(t,t^{\prime})A(t^{\prime})\,\mathrm{d}t^{\prime}\label{eq:noise-corr}
\end{multline}
for $\tau\ge0$. In the latter expression we extended the limits of
integration to infinity, which is possible if we assume $P(t,t^{\prime})=0$
for $t<t^{\prime}$. Equation (\ref{eq:noise-corr}) tells us that
the noise correlation function consists of two parts: a white ($\delta$-correlated)
part and a negative correlation due to neural refractoriness. Intuitively,
since $\delta m_{k}(t)$ and $\delta m_{k+1}(t+\Delta t)$ share the
same number of available neurons $m_{k}(t)$, a positive fluctuation
of the number of spikes in the bin $t$ of the neurons of group $k$
reduces the number of neurons with last spike in bin $k$ more than
on average. Thus, the number of neurons of group $k$ that can still
fire in time bin $t+\Delta t$ is smaller than on average, which explains
the negative correlations (Fig.~\ref{fig:corr-illu}).

\section{Detailed derivation of Eq.~(\ref{eq:pop_dyn_stat})\label{sec:derivation-lin}}

We aim to linearize the population integral $a(t)=\int_{-\infty}^{t}P(t,\hat{t})A(\hat{t})\,\mathrm{d}\hat{t}$,
Eq.~(1), around an equilibrium point $A_{0}$, with small fluctuations
$\Delta A(t)=A(t)-A_{0}$ with a mean of zero. To this end, let us
first note that the hazard function Eq.~(7) then can be written as
\begin{align}
\rho(t,\hat{t}) & =ce^{h_{0}-\vartheta_{0}(t-\hat{t})}e^{\Delta h(t)-\Delta\vartheta(t,\hat{t})}=\rho_{0}(t-\hat{t})e^{\Delta g(t,\hat{t})}.\label{eq:hazard_linearization}
\end{align}
Furthermore, recall that $P(t,\hat{t})=-\nicefrac{\mathrm{d}}{\mathrm{d}t}S(t,\hat{t})$,
where $S(t,\hat{t})=\exp\left(-\int_{\hat{t}}^{t}\rho(s,\hat{t})\mathrm{d}s\right)$.
Expanding to first order in $\Delta A$ yields $S(t,\hat{t})=S_{0}(t-\hat{t})+\Delta S(t,\hat{t})$
and $P(t,\hat{t})=P_{0}(t-\hat{t})-\nicefrac{\mathrm{d}}{\mathrm{d}t}\Delta S(t,\hat{t})$,
where $S_{0}(t)=\exp(-\int_{0}^{t}\rho_{0}(t^{\prime})\mathrm{d}t^{\prime})$
and $P_{0}(t)=-\nicefrac{d}{dt}S_{0}(t)$ define the zeroth-order
terms. Thus, the linearized population integral reads 
\begin{align}
a(t) & =A_{0}+(P_{0}\ast\Delta A)(t)-A_{0}\frac{\mathrm{d}}{\mathrm{d}t}\int_{-\infty}^{t}\Delta S(t,\hat{t})\,\mathrm{d}\hat{t},\label{eq:pop_dyn_proto}
\end{align}
where we used the normalization of $P_{0}(t)$ and the boundary condition
$\Delta S(t,t)=0$.

The perturbation is given by
\begin{align*}
\Delta S(t,\hat{t}) & =\int_{-\infty}^{\infty}\left.\frac{\delta S(t,\hat{t})}{\delta\Delta g(s,\hat{t})}\right|_{\Delta g=0}\cdot\Delta g(s,\hat{t})\,\mathrm{d}s.
\end{align*}
The functional derivative at $\Delta g=0$ reads 
\begin{align*}
\left.\frac{\delta S(t,\hat{t})}{\delta\Delta g(s,\hat{t})}\right|_{\Delta g=0} & =\left.\frac{\delta\exp\left(-\int_{\hat{t}}^{t}\rho_{0}(t^{\prime}-\hat{t})e^{\Delta g(t^{\prime},\hat{t})}dt^{\prime}\right)}{\delta\Delta g(s,\hat{t})}\right|_{\Delta g=0}\\
 & =-\theta(t-s)\theta(s-\hat{t})S_{0}(t,\hat{t})\rho_{0}(s-\hat{t}).
\end{align*}

We insert this expression to compute the integral in (\ref{eq:pop_dyn_proto})
\begin{multline*}
-\int_{-\infty}^{t}\Delta S(t,\hat{t})\mathrm{d}\hat{t}=\int_{-\infty}^{t}\int_{-\infty}^{\infty}\theta(s-\hat{t})\theta(t-s)\\
\times S_{0}(t-\hat{t})\rho_{0}(\underbrace{s-\hat{t}}_{x(\hat{t})})\Delta g(s,\hat{t})\,\mathrm{d}s\,\mathrm{d}\hat{t}\\
=\int_{-\infty}^{\infty}\theta(t-s)\int_{0}^{\infty}\theta(x)S_{0}(t-s+x)\rho_{0}(x)\Delta g(s,s-x)\mathrm{d}x\mathrm{d}s\\
\overset{[8]}{=}\int_{-\infty}^{\infty}\Delta g(s)\underbrace{\theta(t-s)\int_{0}^{\infty}\theta(x)S_{0}(t-s+x)\rho_{0}(x)\,\mathrm{d}x}_{\mathcal{L}(t-s)}\,\mathrm{d}s\,.
\end{multline*}
In the last step, we have used the approximation Eq.~(8), $\Delta g(t,\hat{t})=\Delta g(t)=[(\kappa J_{\mathrm{s}}-\gamma)\ast\Delta A](t)$,
which has no dependence on the time of the last spike. Hence (\ref{eq:pop_dyn_proto})
becomes 
\begin{align}
a(t) & =A_{0}+(P_{0}\ast\Delta A)(t)+A_{0}\frac{\mathrm{d}}{\mathrm{d}t}(\mathcal{L}\ast\Delta g)(t).\label{eq:pop_dyn_mean}
\end{align}
This linearized equation for the mean activity is valid for any small
$\Delta A$ and $\Delta g$ in the past, also if due to finite size
fluctuations. Eqs.~(\ref{eq:hazard_linearization})-(\ref{eq:pop_dyn_mean})
generalize to additional time-dependent inputs $\Delta I(t)$ by extending
the definition of $\Delta g$ to $\Delta g(t)=[(\kappa J_{\mathrm{s}}-\gamma)\ast\Delta A+\kappa\ast\Delta I](t)$.

Furthermore, the noise correlation function Eq.~(\ref{eq:noise-corr})
becomes to leading order 
\begin{multline}
\langle\xi(t+\tau)\xi(t)\rangle=\delta(\tau)-\int_{-\infty}^{t}P_{0}(t+\tau-\hat{t})P_{0}(t-\hat{t})\,\mathrm{d}\hat{t},\label{eq:pop_corr_stat}
\end{multline}
with $\sqrt{A_{0}/N}\xi(t)=\delta A(t)$. Eqs.~(\ref{eq:pop_dyn_mean})
and (\ref{eq:pop_corr_stat}) are equivalent to Eq.~(\ref{eq:pop_dyn_stat}).

\bibliographystyle{apsrev4-1}
\bibliography{../bib/deger_jabref}

\begin{thebibliography}{65}%
\makeatletter
\providecommand \@ifxundefined [1]{%
 \@ifx{#1\undefined}
}%
\providecommand \@ifnum [1]{%
 \ifnum #1\expandafter \@firstoftwo
 \else \expandafter \@secondoftwo
 \fi
}%
\providecommand \@ifx [1]{%
 \ifx #1\expandafter \@firstoftwo
 \else \expandafter \@secondoftwo
 \fi
}%
\providecommand \natexlab [1]{#1}%
\providecommand \enquote  [1]{``#1''}%
\providecommand \bibnamefont  [1]{#1}%
\providecommand \bibfnamefont [1]{#1}%
\providecommand \citenamefont [1]{#1}%
\providecommand \href@noop [0]{\@secondoftwo}%
\providecommand \href [0]{\begingroup \@sanitize@url \@href}%
\providecommand \@href[1]{\@@startlink{#1}\@@href}%
\providecommand \@@href[1]{\endgroup#1\@@endlink}%
\providecommand \@sanitize@url [0]{\catcode `\\12\catcode `\$12\catcode
  `\&12\catcode `\#12\catcode `\^12\catcode `\_12\catcode `\%12\relax}%
\providecommand \@@startlink[1]{}%
\providecommand \@@endlink[0]{}%
\providecommand \url  [0]{\begingroup\@sanitize@url \@url }%
\providecommand \@url [1]{\endgroup\@href {#1}{\urlprefix }}%
\providecommand \urlprefix  [0]{URL }%
\providecommand \Eprint [0]{\href }%
\providecommand \doibase [0]{http://dx.doi.org/}%
\providecommand \selectlanguage [0]{\@gobble}%
\providecommand \bibinfo  [0]{\@secondoftwo}%
\providecommand \bibfield  [0]{\@secondoftwo}%
\providecommand \translation [1]{[#1]}%
\providecommand \BibitemOpen [0]{}%
\providecommand \bibitemStop [0]{}%
\providecommand \bibitemNoStop [0]{.\EOS\space}%
\providecommand \EOS [0]{\spacefactor3000\relax}%
\providecommand \BibitemShut  [1]{\csname bibitem#1\endcsname}%
\let\auto@bib@innerbib\@empty
\bibitem [{\citenamefont {Klein}\ and\ \citenamefont
  {Shinoda}(2008)}]{Klein2008}%
  \BibitemOpen
  \bibfield  {author} {\bibinfo {author} {\bibfnamefont {M.~L.}\ \bibnamefont
  {Klein}}\ and\ \bibinfo {author} {\bibfnamefont {W.}~\bibnamefont
  {Shinoda}},\ }\href {\doibase 10.1126/science.1157834} {\bibfield  {journal}
  {\bibinfo  {journal} {Science}\ }\textbf {\bibinfo {volume} {321}},\ \bibinfo
  {pages} {798} (\bibinfo {year} {2008})}\BibitemShut {NoStop}%
\bibitem [{\citenamefont {Ayton}\ \emph {et~al.}(2007)\citenamefont {Ayton},
  \citenamefont {Noid},\ and\ \citenamefont {Voth}}]{Ayton2007}%
  \BibitemOpen
  \bibfield  {author} {\bibinfo {author} {\bibfnamefont {G.~S.}\ \bibnamefont
  {Ayton}}, \bibinfo {author} {\bibfnamefont {W.~G.}\ \bibnamefont {Noid}}, \
  and\ \bibinfo {author} {\bibfnamefont {G.~A.}\ \bibnamefont {Voth}},\ }\href
  {\doibase 10.1016/j.sbi.2007.03.004} {\bibfield  {journal} {\bibinfo
  {journal} {Curr Opin Struct Biol}\ }\textbf {\bibinfo {volume} {17}},\
  \bibinfo {pages} {192} (\bibinfo {year} {2007})}\BibitemShut {NoStop}%
\bibitem [{\citenamefont {Peter}\ and\ \citenamefont
  {Kremer}(2009)}]{Peter2009}%
  \BibitemOpen
  \bibfield  {author} {\bibinfo {author} {\bibfnamefont {C.}~\bibnamefont
  {Peter}}\ and\ \bibinfo {author} {\bibfnamefont {K.}~\bibnamefont {Kremer}},\
  }\href {\doibase 10.1039/b912027k} {\bibfield  {journal} {\bibinfo  {journal}
  {Soft Matter}\ }\textbf {\bibinfo {volume} {5}},\ \bibinfo {pages} {4357}
  (\bibinfo {year} {2009})}\BibitemShut {NoStop}%
\bibitem [{\citenamefont {Pielke}(2002)}]{Pielke2002}%
  \BibitemOpen
  \bibfield  {author} {\bibinfo {author} {\bibfnamefont {R.~A.}\ \bibnamefont
  {Pielke}},\ }\href@noop {} {\emph {\bibinfo {title} {Mesoscale Meteorological
  Modeling}}}\ (\bibinfo  {publisher} {Academic Press},\ \bibinfo {year}
  {2002})\BibitemShut {NoStop}%
\bibitem [{\citenamefont {Bena\"{i}m}\ and\ \citenamefont
  {Le\;Boudec}(2008)}]{Benaim2008}%
  \BibitemOpen
  \bibfield  {author} {\bibinfo {author} {\bibfnamefont {M.}~\bibnamefont
  {Bena\"{i}m}}\ and\ \bibinfo {author} {\bibfnamefont {J.-Y.}\ \bibnamefont
  {Le\;Boudec}},\ }\href {\doibase 10.1016/j.peva.2008.03.005} {\bibfield
  {journal} {\bibinfo  {journal} {Perform Evaluation}\ }\textbf {\bibinfo
  {volume} {65}},\ \bibinfo {pages} {823} (\bibinfo {year} {2008})}\BibitemShut
  {NoStop}%
\bibitem [{\citenamefont {Gerstner}\ \emph {et~al.}(2012)\citenamefont
  {Gerstner}, \citenamefont {Sprekeler},\ and\ \citenamefont
  {Deco}}]{Gerstner2012}%
  \BibitemOpen
  \bibfield  {author} {\bibinfo {author} {\bibfnamefont {W.}~\bibnamefont
  {Gerstner}}, \bibinfo {author} {\bibfnamefont {H.}~\bibnamefont {Sprekeler}},
  \ and\ \bibinfo {author} {\bibfnamefont {G.}~\bibnamefont {Deco}},\ }\href
  {\doibase 10.1126/science.1227356} {\bibfield  {journal} {\bibinfo  {journal}
  {Science}\ }\textbf {\bibinfo {volume} {338}},\ \bibinfo {pages} {60}
  (\bibinfo {year} {2012})}\BibitemShut {NoStop}%
\bibitem [{\citenamefont {Lefort}\ \emph {et~al.}(2009)\citenamefont {Lefort},
  \citenamefont {Tomm}, \citenamefont {Sarria},\ and\ \citenamefont
  {Petersen}}]{Lefort2009}%
  \BibitemOpen
  \bibfield  {author} {\bibinfo {author} {\bibfnamefont {S.}~\bibnamefont
  {Lefort}}, \bibinfo {author} {\bibfnamefont {C.}~\bibnamefont {Tomm}},
  \bibinfo {author} {\bibfnamefont {J.-C.~F.}\ \bibnamefont {Sarria}}, \ and\
  \bibinfo {author} {\bibfnamefont {C.~C.~H.}\ \bibnamefont {Petersen}},\
  }\href {\doibase 10.1016/j.neuron.2008.12.020} {\bibfield  {journal}
  {\bibinfo  {journal} {Neuron}\ }\textbf {\bibinfo {volume} {61}},\ \bibinfo
  {pages} {301} (\bibinfo {year} {2009})}\BibitemShut {NoStop}%
\bibitem [{\citenamefont {Hubel}\ and\ \citenamefont
  {Wiesel}(1962)}]{Hubel1962}%
  \BibitemOpen
  \bibfield  {author} {\bibinfo {author} {\bibfnamefont {D.~H.}\ \bibnamefont
  {Hubel}}\ and\ \bibinfo {author} {\bibfnamefont {T.~N.}\ \bibnamefont
  {Wiesel}},\ }\href@noop {} {\bibfield  {journal} {\bibinfo  {journal} {J
  Physiol}\ }\textbf {\bibinfo {volume} {160}},\ \bibinfo {pages} {106}
  (\bibinfo {year} {1962})}\BibitemShut {NoStop}%
\bibitem [{\citenamefont {Mensi}\ \emph {et~al.}(2012)\citenamefont {Mensi},
  \citenamefont {Naud}, \citenamefont {Pozzorini}, \citenamefont {Avermann},
  \citenamefont {Petersen},\ and\ \citenamefont {Gerstner}}]{Mensi2012}%
  \BibitemOpen
  \bibfield  {author} {\bibinfo {author} {\bibfnamefont {S.}~\bibnamefont
  {Mensi}}, \bibinfo {author} {\bibfnamefont {R.}~\bibnamefont {Naud}},
  \bibinfo {author} {\bibfnamefont {C.}~\bibnamefont {Pozzorini}}, \bibinfo
  {author} {\bibfnamefont {M.}~\bibnamefont {Avermann}}, \bibinfo {author}
  {\bibfnamefont {C.~C.~H.}\ \bibnamefont {Petersen}}, \ and\ \bibinfo {author}
  {\bibfnamefont {W.}~\bibnamefont {Gerstner}},\ }\href {\doibase
  10.1152/jn.00408.2011} {\bibfield  {journal} {\bibinfo  {journal} {J
  Neurophysiol}\ }\textbf {\bibinfo {volume} {107}},\ \bibinfo {pages} {1756}
  (\bibinfo {year} {2012})}\BibitemShut {NoStop}%
\bibitem [{\citenamefont {Brunel}\ and\ \citenamefont
  {Hakim}(1999)}]{Brunel1999}%
  \BibitemOpen
  \bibfield  {author} {\bibinfo {author} {\bibfnamefont {N.}~\bibnamefont
  {Brunel}}\ and\ \bibinfo {author} {\bibfnamefont {V.}~\bibnamefont {Hakim}},\
  }\href {\doibase 10.1162/089976699300016179} {\bibfield  {journal} {\bibinfo
  {journal} {Neural Comput}\ }\textbf {\bibinfo {volume} {11}},\ \bibinfo
  {pages} {1621} (\bibinfo {year} {1999})}\BibitemShut {NoStop}%
\bibitem [{\citenamefont {Gerstner}(2000)}]{Gerstner2000}%
  \BibitemOpen
  \bibfield  {author} {\bibinfo {author} {\bibfnamefont {W.}~\bibnamefont
  {Gerstner}},\ }\href {\doibase 10.1162/089976600300015899} {\bibfield
  {journal} {\bibinfo  {journal} {Neural Comput}\ }\textbf {\bibinfo {volume}
  {12}},\ \bibinfo {pages} {43} (\bibinfo {year} {2000})}\BibitemShut {NoStop}%
\bibitem [{\citenamefont {Pillow}\ \emph {et~al.}(2008)\citenamefont {Pillow},
  \citenamefont {Shlens}, \citenamefont {Paninski}, \citenamefont {Sher},
  \citenamefont {Litke}, \citenamefont {Chichilnisky},\ and\ \citenamefont
  {Simoncelli}}]{Pillow2008}%
  \BibitemOpen
  \bibfield  {author} {\bibinfo {author} {\bibfnamefont {J.~W.}\ \bibnamefont
  {Pillow}}, \bibinfo {author} {\bibfnamefont {J.}~\bibnamefont {Shlens}},
  \bibinfo {author} {\bibfnamefont {L.}~\bibnamefont {Paninski}}, \bibinfo
  {author} {\bibfnamefont {A.}~\bibnamefont {Sher}}, \bibinfo {author}
  {\bibfnamefont {A.~M.}\ \bibnamefont {Litke}}, \bibinfo {author}
  {\bibfnamefont {E.~J.}\ \bibnamefont {Chichilnisky}}, \ and\ \bibinfo
  {author} {\bibfnamefont {E.~P.}\ \bibnamefont {Simoncelli}},\ }\href
  {\doibase 10.1038/nature07140} {\bibfield  {journal} {\bibinfo  {journal}
  {Nature}\ }\textbf {\bibinfo {volume} {454}},\ \bibinfo {pages} {995}
  (\bibinfo {year} {2008})}\BibitemShut {NoStop}%
\bibitem [{\citenamefont {Frost}\ and\ \citenamefont
  {Melamed}(1994)}]{Frost1994}%
  \BibitemOpen
  \bibfield  {author} {\bibinfo {author} {\bibfnamefont {V.}~\bibnamefont
  {Frost}}\ and\ \bibinfo {author} {\bibfnamefont {B.}~\bibnamefont
  {Melamed}},\ }\href {\doibase 10.1109/35.267444} {\bibfield  {journal}
  {\bibinfo  {journal} {IEEE C M}\ }\textbf {\bibinfo {volume} {32}},\ \bibinfo
  {pages} {70} (\bibinfo {year} {1994})}\BibitemShut {NoStop}%
\bibitem [{\citenamefont {Mirollo}\ and\ \citenamefont
  {Strogatz}(1990)}]{Mirollo90}%
  \BibitemOpen
  \bibfield  {author} {\bibinfo {author} {\bibfnamefont {R.~E.}\ \bibnamefont
  {Mirollo}}\ and\ \bibinfo {author} {\bibfnamefont {S.~H.}\ \bibnamefont
  {Strogatz}},\ }\href {\doibase 10.1137/0150098} {\bibfield  {journal}
  {\bibinfo  {journal} {SIAM J Appl Math}\ }\textbf {\bibinfo {volume} {50}},\
  \bibinfo {pages} {1645} (\bibinfo {year} {1990})}\BibitemShut {NoStop}%
\bibitem [{\citenamefont {Mattia}\ and\ \citenamefont
  {Del~Giudice}(2002)}]{Mattia2002}%
  \BibitemOpen
  \bibfield  {author} {\bibinfo {author} {\bibfnamefont {M.}~\bibnamefont
  {Mattia}}\ and\ \bibinfo {author} {\bibfnamefont {P.}~\bibnamefont
  {Del~Giudice}},\ }\href {\doibase 10.1103/PhysRevE.66.051917} {\bibfield
  {journal} {\bibinfo  {journal} {Phys Rev E}\ }\textbf {\bibinfo {volume}
  {66}},\ \bibinfo {pages} {051917} (\bibinfo {year} {2002})}\BibitemShut
  {NoStop}%
\bibitem [{\citenamefont {Buice}\ and\ \citenamefont {Chow}(2013)}]{Buice2013}%
  \BibitemOpen
  \bibfield  {author} {\bibinfo {author} {\bibfnamefont {M.~A.}\ \bibnamefont
  {Buice}}\ and\ \bibinfo {author} {\bibfnamefont {C.~C.}\ \bibnamefont
  {Chow}},\ }\href {\doibase 10.1371/journal.pcbi.1002872} {\bibfield
  {journal} {\bibinfo  {journal} {PLoS Comput Biol}\ }\textbf {\bibinfo
  {volume} {9}},\ \bibinfo {pages} {e1002872} (\bibinfo {year}
  {2013})}\BibitemShut {NoStop}%
\bibitem [{\citenamefont {Toyoizumi}\ \emph {et~al.}(2009)\citenamefont
  {Toyoizumi}, \citenamefont {Rad},\ and\ \citenamefont
  {Paninski}}]{Toyoizumi2009}%
  \BibitemOpen
  \bibfield  {author} {\bibinfo {author} {\bibfnamefont {T.}~\bibnamefont
  {Toyoizumi}}, \bibinfo {author} {\bibfnamefont {K.~R.}\ \bibnamefont {Rad}},
  \ and\ \bibinfo {author} {\bibfnamefont {L.}~\bibnamefont {Paninski}},\
  }\href {\doibase 10.1162/neco.2008.04-08-757} {\bibfield  {journal} {\bibinfo
   {journal} {Neural Comput}\ }\textbf {\bibinfo {volume} {21}},\ \bibinfo
  {pages} {1203} (\bibinfo {year} {2009})}\BibitemShut {NoStop}%
\bibitem [{\citenamefont {Hawkes}(1971)}]{Hawkes1971a}%
  \BibitemOpen
  \bibfield  {author} {\bibinfo {author} {\bibfnamefont {A.~G.}\ \bibnamefont
  {Hawkes}},\ }\href@noop {} {\bibfield  {journal} {\bibinfo  {journal}
  {Journal of the Royal Statistical Society. Series B (Methodological)}\
  }\textbf {\bibinfo {volume} {33}},\ \bibinfo {pages} {438} (\bibinfo {year}
  {1971})}\BibitemShut {NoStop}%
\bibitem [{\citenamefont {Pernice}\ \emph {et~al.}(2012)\citenamefont
  {Pernice}, \citenamefont {Staude}, \citenamefont {Cardanobile},\ and\
  \citenamefont {Rotter}}]{Pernice2012}%
  \BibitemOpen
  \bibfield  {author} {\bibinfo {author} {\bibfnamefont {V.}~\bibnamefont
  {Pernice}}, \bibinfo {author} {\bibfnamefont {B.}~\bibnamefont {Staude}},
  \bibinfo {author} {\bibfnamefont {S.}~\bibnamefont {Cardanobile}}, \ and\
  \bibinfo {author} {\bibfnamefont {S.}~\bibnamefont {Rotter}},\ }\href
  {\doibase 10.1103/PhysRevE.85.031916} {\bibfield  {journal} {\bibinfo
  {journal} {Phys Rev E}\ }\textbf {\bibinfo {volume} {85}},\ \bibinfo {pages}
  {031916} (\bibinfo {year} {2012})}\BibitemShut {NoStop}%
\bibitem [{\citenamefont {Helias}\ \emph {et~al.}(2013)\citenamefont {Helias},
  \citenamefont {Tetzlaff},\ and\ \citenamefont {Diesmann}}]{Helias2013}%
  \BibitemOpen
  \bibfield  {author} {\bibinfo {author} {\bibfnamefont {M.}~\bibnamefont
  {Helias}}, \bibinfo {author} {\bibfnamefont {T.}~\bibnamefont {Tetzlaff}}, \
  and\ \bibinfo {author} {\bibfnamefont {M.}~\bibnamefont {Diesmann}},\ }\href
  {\doibase 10.1088/1367-2630/15/2/023002} {\bibfield  {journal} {\bibinfo
  {journal} {New J Phys}\ }\textbf {\bibinfo {volume} {15}},\ \bibinfo {pages}
  {023002} (\bibinfo {year} {2013})}\BibitemShut {NoStop}%
\bibitem [{\citenamefont {Soula}\ and\ \citenamefont {Chow}(2007)}]{Soula2007}%
  \BibitemOpen
  \bibfield  {author} {\bibinfo {author} {\bibfnamefont {H.}~\bibnamefont
  {Soula}}\ and\ \bibinfo {author} {\bibfnamefont {C.~C.}\ \bibnamefont
  {Chow}},\ }\href {\doibase 10.1162/neco.2007.19.12.3262} {\bibfield
  {journal} {\bibinfo  {journal} {Neural Comput}\ }\textbf {\bibinfo {volume}
  {19}},\ \bibinfo {pages} {3262} (\bibinfo {year} {2007})}\BibitemShut
  {NoStop}%
\bibitem [{\citenamefont {Buice}\ and\ \citenamefont
  {Cowan}(2007)}]{Buice2007}%
  \BibitemOpen
  \bibfield  {author} {\bibinfo {author} {\bibfnamefont {M.~A.}\ \bibnamefont
  {Buice}}\ and\ \bibinfo {author} {\bibfnamefont {J.~D.}\ \bibnamefont
  {Cowan}},\ }\href {\doibase 10.1103/PhysRevE.75.051919} {\bibfield  {journal}
  {\bibinfo  {journal} {Phys Rev E}\ }\textbf {\bibinfo {volume} {75}},\
  \bibinfo {pages} {051919} (\bibinfo {year} {2007})}\BibitemShut {NoStop}%
\bibitem [{\citenamefont {Bressloff}(2009)}]{Bressloff2009}%
  \BibitemOpen
  \bibfield  {author} {\bibinfo {author} {\bibfnamefont {P.~C.}\ \bibnamefont
  {Bressloff}},\ }\href {\doibase 10.1137/090756971} {\bibfield  {journal}
  {\bibinfo  {journal} {SIAM J Appl Math}\ }\textbf {\bibinfo {volume} {70}},\
  \bibinfo {pages} {1488} (\bibinfo {year} {2009})}\BibitemShut {NoStop}%
\bibitem [{\citenamefont {Benayoun}\ \emph {et~al.}(2010)\citenamefont
  {Benayoun}, \citenamefont {Cowan}, \citenamefont {van Drongelen},\ and\
  \citenamefont {Wallace}}]{Benayoun2010a}%
  \BibitemOpen
  \bibfield  {author} {\bibinfo {author} {\bibfnamefont {M.}~\bibnamefont
  {Benayoun}}, \bibinfo {author} {\bibfnamefont {J.~D.}\ \bibnamefont {Cowan}},
  \bibinfo {author} {\bibfnamefont {W.}~\bibnamefont {van Drongelen}}, \ and\
  \bibinfo {author} {\bibfnamefont {E.}~\bibnamefont {Wallace}},\ }\href
  {\doibase 10.1371/journal.pcbi.1000846} {\bibfield  {journal} {\bibinfo
  {journal} {PLoS Comput Biol}\ }\textbf {\bibinfo {volume} {6}},\ \bibinfo
  {pages} {e1000846} (\bibinfo {year} {2010})}\BibitemShut {NoStop}%
\bibitem [{\citenamefont {Touboul}\ and\ \citenamefont
  {Ermentrout}(2011)}]{Touboul2011}%
  \BibitemOpen
  \bibfield  {author} {\bibinfo {author} {\bibfnamefont {J.~D.}\ \bibnamefont
  {Touboul}}\ and\ \bibinfo {author} {\bibfnamefont {G.~B.}\ \bibnamefont
  {Ermentrout}},\ }\href {\doibase 10.1007/s10827-011-0320-5} {\bibfield
  {journal} {\bibinfo  {journal} {J Comput Neurosci}\ }\textbf {\bibinfo
  {volume} {31}},\ \bibinfo {pages} {453} (\bibinfo {year} {2011})}\BibitemShut
  {NoStop}%
\bibitem [{\citenamefont {Dumont}\ \emph {et~al.}(2014)\citenamefont {Dumont},
  \citenamefont {Northoff},\ and\ \citenamefont {Longtin}}]{Dumont2014}%
  \BibitemOpen
  \bibfield  {author} {\bibinfo {author} {\bibfnamefont {G.}~\bibnamefont
  {Dumont}}, \bibinfo {author} {\bibfnamefont {G.}~\bibnamefont {Northoff}}, \
  and\ \bibinfo {author} {\bibfnamefont {A.}~\bibnamefont {Longtin}},\ }\href
  {\doibase 10.1103/PhysRevE.90.012702} {\bibfield  {journal} {\bibinfo
  {journal} {Phys Rev E}\ }\textbf {\bibinfo {volume} {90}},\ \bibinfo {pages}
  {012702} (\bibinfo {year} {2014})}\BibitemShut {NoStop}%
\bibitem [{\citenamefont {Lagzi}\ and\ \citenamefont
  {Rotter}(2014)}]{Lagzi2014}%
  \BibitemOpen
  \bibfield  {author} {\bibinfo {author} {\bibfnamefont {F.}~\bibnamefont
  {Lagzi}}\ and\ \bibinfo {author} {\bibfnamefont {S.}~\bibnamefont {Rotter}},\
  }\href {\doibase 10.3389/fncom.2014.00142} {\bibfield  {journal} {\bibinfo
  {journal} {Front Comput Neurosci}\ }\textbf {\bibinfo {volume} {8}},\
  \bibinfo {pages} {142} (\bibinfo {year} {2014})}\BibitemShut {NoStop}%
\bibitem [{\citenamefont {Lindner}\ \emph {et~al.}(2005)\citenamefont
  {Lindner}, \citenamefont {Doiron},\ and\ \citenamefont
  {Longtin}}]{Lindner2005}%
  \BibitemOpen
  \bibfield  {author} {\bibinfo {author} {\bibfnamefont {B.}~\bibnamefont
  {Lindner}}, \bibinfo {author} {\bibfnamefont {B.}~\bibnamefont {Doiron}}, \
  and\ \bibinfo {author} {\bibfnamefont {A.}~\bibnamefont {Longtin}},\ }\href
  {\doibase 10.1103/PhysRevE.72.061919} {\bibfield  {journal} {\bibinfo
  {journal} {Phys Rev E}\ }\textbf {\bibinfo {volume} {72}},\ \bibinfo {pages}
  {061919} (\bibinfo {year} {2005})}\BibitemShut {NoStop}%
\bibitem [{\citenamefont {\'Avila~\AA{}kerberg}\ and\ \citenamefont
  {Chacron}(2009)}]{Akerberg2009}%
  \BibitemOpen
  \bibfield  {author} {\bibinfo {author} {\bibfnamefont {O.}~\bibnamefont
  {\'Avila~\AA{}kerberg}}\ and\ \bibinfo {author} {\bibfnamefont {M.~J.}\
  \bibnamefont {Chacron}},\ }\href {\doibase 10.1103/PhysRevE.79.011914}
  {\bibfield  {journal} {\bibinfo  {journal} {Phys Rev E}\ }\textbf {\bibinfo
  {volume} {79}},\ \bibinfo {pages} {011914} (\bibinfo {year}
  {2009})}\BibitemShut {NoStop}%
\bibitem [{\citenamefont {Trousdale}\ \emph {et~al.}(2012)\citenamefont
  {Trousdale}, \citenamefont {Hu}, \citenamefont {Shea-Brown},\ and\
  \citenamefont {Josi\'{c}}}]{Trousdale2012}%
  \BibitemOpen
  \bibfield  {author} {\bibinfo {author} {\bibfnamefont {J.}~\bibnamefont
  {Trousdale}}, \bibinfo {author} {\bibfnamefont {Y.}~\bibnamefont {Hu}},
  \bibinfo {author} {\bibfnamefont {E.}~\bibnamefont {Shea-Brown}}, \ and\
  \bibinfo {author} {\bibfnamefont {K.}~\bibnamefont {Josi\'{c}}},\ }\href
  {\doibase 10.1371/journal.pcbi.1002408} {\bibfield  {journal} {\bibinfo
  {journal} {PLoS Comput Biol}\ }\textbf {\bibinfo {volume} {8}},\ \bibinfo
  {pages} {e1002408} (\bibinfo {year} {2012})}\BibitemShut {NoStop}%
\bibitem [{\citenamefont {Benda}\ and\ \citenamefont
  {Herz}(2003)}]{Benda2003c}%
  \BibitemOpen
  \bibfield  {author} {\bibinfo {author} {\bibfnamefont {J.}~\bibnamefont
  {Benda}}\ and\ \bibinfo {author} {\bibfnamefont {A.~V.~M.}\ \bibnamefont
  {Herz}},\ }\href {\doibase 10.1162/089976603322385063} {\bibfield  {journal}
  {\bibinfo  {journal} {Neural Comput}\ }\textbf {\bibinfo {volume} {15}},\
  \bibinfo {pages} {2523} (\bibinfo {year} {2003})}\BibitemShut {NoStop}%
\bibitem [{\citenamefont {Lundstrom}\ \emph {et~al.}(2008)\citenamefont
  {Lundstrom}, \citenamefont {Higgs}, \citenamefont {Spain},\ and\
  \citenamefont {Fairhall}}]{Lundstrom2008}%
  \BibitemOpen
  \bibfield  {author} {\bibinfo {author} {\bibfnamefont {B.~N.}\ \bibnamefont
  {Lundstrom}}, \bibinfo {author} {\bibfnamefont {M.~H.}\ \bibnamefont
  {Higgs}}, \bibinfo {author} {\bibfnamefont {W.~J.}\ \bibnamefont {Spain}}, \
  and\ \bibinfo {author} {\bibfnamefont {A.~L.}\ \bibnamefont {Fairhall}},\
  }\href {\doibase 10.1038/nn.2212} {\bibfield  {journal} {\bibinfo  {journal}
  {Nat Neurosci}\ }\textbf {\bibinfo {volume} {11}},\ \bibinfo {pages} {1335}
  (\bibinfo {year} {2008})}\BibitemShut {NoStop}%
\bibitem [{\citenamefont {Pozzorini}\ \emph {et~al.}(2013)\citenamefont
  {Pozzorini}, \citenamefont {Naud}, \citenamefont {Mensi},\ and\ \citenamefont
  {Gerstner}}]{Pozzorini2013}%
  \BibitemOpen
  \bibfield  {author} {\bibinfo {author} {\bibfnamefont {C.}~\bibnamefont
  {Pozzorini}}, \bibinfo {author} {\bibfnamefont {R.}~\bibnamefont {Naud}},
  \bibinfo {author} {\bibfnamefont {S.}~\bibnamefont {Mensi}}, \ and\ \bibinfo
  {author} {\bibfnamefont {W.}~\bibnamefont {Gerstner}},\ }\href {\doibase
  10.1038/nn.3431} {\bibfield  {journal} {\bibinfo  {journal} {Nat Neurosci}\
  }\textbf {\bibinfo {volume} {16}},\ \bibinfo {pages} {942} (\bibinfo {year}
  {2013})}\BibitemShut {NoStop}%
\bibitem [{\citenamefont {Farkhooi}\ \emph {et~al.}(2013)\citenamefont
  {Farkhooi}, \citenamefont {Froese}, \citenamefont {Muller}, \citenamefont
  {Menzel},\ and\ \citenamefont {Nawrot}}]{Farkhooi2013}%
  \BibitemOpen
  \bibfield  {author} {\bibinfo {author} {\bibfnamefont {F.}~\bibnamefont
  {Farkhooi}}, \bibinfo {author} {\bibfnamefont {A.}~\bibnamefont {Froese}},
  \bibinfo {author} {\bibfnamefont {E.}~\bibnamefont {Muller}}, \bibinfo
  {author} {\bibfnamefont {R.}~\bibnamefont {Menzel}}, \ and\ \bibinfo {author}
  {\bibfnamefont {M.~P.}\ \bibnamefont {Nawrot}},\ }\href {\doibase
  10.1371/journal.pcbi.1003251} {\bibfield  {journal} {\bibinfo  {journal}
  {PLoS Comput Biol}\ }\textbf {\bibinfo {volume} {9}},\ \bibinfo {pages}
  {e1003251} (\bibinfo {year} {2013})}\BibitemShut {NoStop}%
\bibitem [{\citenamefont {Prescott}\ and\ \citenamefont
  {Sejnowski}(2008)}]{Prescott2008}%
  \BibitemOpen
  \bibfield  {author} {\bibinfo {author} {\bibfnamefont {S.~A.}\ \bibnamefont
  {Prescott}}\ and\ \bibinfo {author} {\bibfnamefont {T.~J.}\ \bibnamefont
  {Sejnowski}},\ }\href {\doibase 10.1523/JNEUROSCI.1792-08.2008} {\bibfield
  {journal} {\bibinfo  {journal} {J Neurosci}\ }\textbf {\bibinfo {volume}
  {28}},\ \bibinfo {pages} {13649} (\bibinfo {year} {2008})}\BibitemShut
  {NoStop}%
\bibitem [{\citenamefont {Schwalger}(2013)}]{Schwalger2013a}%
  \BibitemOpen
  \bibfield  {author} {\bibinfo {author} {\bibfnamefont {T.}~\bibnamefont
  {Schwalger}},\ }\emph {\bibinfo {title} {The interspike-interval statistics
  of non-renewal neuron models}},\ \href
  {http://edoc.hu-berlin.de/docviews/abstract.php?id=40328} {Ph.D. thesis},\
  \bibinfo  {school} {Humboldt-Universit\"at zu Berlin} (\bibinfo {year}
  {2013})\BibitemShut {NoStop}%
\bibitem [{\citenamefont {Schwalger}\ and\ \citenamefont
  {Lindner}(2013)}]{Schwalger2013}%
  \BibitemOpen
  \bibfield  {author} {\bibinfo {author} {\bibfnamefont {T.}~\bibnamefont
  {Schwalger}}\ and\ \bibinfo {author} {\bibfnamefont {B.}~\bibnamefont
  {Lindner}},\ }\href {\doibase 10.3389/fncom.2013.00164} {\bibfield  {journal}
  {\bibinfo  {journal} {Front Comput Neurosci}\ }\textbf {\bibinfo {volume}
  {7}},\ \bibinfo {pages} {164} (\bibinfo {year} {2013})}\BibitemShut {NoStop}%
\bibitem [{\citenamefont {Naud}\ and\ \citenamefont
  {Gerstner}(2012)}]{Naud2012}%
  \BibitemOpen
  \bibfield  {author} {\bibinfo {author} {\bibfnamefont {R.}~\bibnamefont
  {Naud}}\ and\ \bibinfo {author} {\bibfnamefont {W.}~\bibnamefont
  {Gerstner}},\ }\href {\doibase 10.1371/journal.pcbi.1002711} {\bibfield
  {journal} {\bibinfo  {journal} {PLoS Comput Biol}\ }\textbf {\bibinfo
  {volume} {8}},\ \bibinfo {pages} {e1002711} (\bibinfo {year}
  {2012})}\BibitemShut {NoStop}%
\bibitem [{\citenamefont {Cox}\ and\ \citenamefont {Isham}(1980)}]{Cox1980}%
  \BibitemOpen
  \bibfield  {author} {\bibinfo {author} {\bibfnamefont {D.}~\bibnamefont
  {Cox}}\ and\ \bibinfo {author} {\bibfnamefont {V.}~\bibnamefont {Isham}},\
  }\href@noop {} {\emph {\bibinfo {title} {Point processes}}}\ (\bibinfo
  {publisher} {Chapman and Hall},\ \bibinfo {year} {1980})\BibitemShut
  {NoStop}%
\bibitem [{\citenamefont {Wilson}\ and\ \citenamefont
  {Cowan}(1972)}]{Wilson1972}%
  \BibitemOpen
  \bibfield  {author} {\bibinfo {author} {\bibfnamefont {H.~R.}\ \bibnamefont
  {Wilson}}\ and\ \bibinfo {author} {\bibfnamefont {J.~D.}\ \bibnamefont
  {Cowan}},\ }\href {\doibase 10.1016/S0006-3495(72)86068-5} {\bibfield
  {journal} {\bibinfo  {journal} {Biophys J}\ }\textbf {\bibinfo {volume}
  {12}},\ \bibinfo {pages} {1} (\bibinfo {year} {1972})}\BibitemShut {NoStop}%
\bibitem [{\citenamefont {Deger}\ \emph {et~al.}(2010)\citenamefont {Deger},
  \citenamefont {Helias}, \citenamefont {Cardanobile}, \citenamefont {Atay},\
  and\ \citenamefont {Rotter}}]{Deger2010}%
  \BibitemOpen
  \bibfield  {author} {\bibinfo {author} {\bibfnamefont {M.}~\bibnamefont
  {Deger}}, \bibinfo {author} {\bibfnamefont {M.}~\bibnamefont {Helias}},
  \bibinfo {author} {\bibfnamefont {S.}~\bibnamefont {Cardanobile}}, \bibinfo
  {author} {\bibfnamefont {F.~M.}\ \bibnamefont {Atay}}, \ and\ \bibinfo
  {author} {\bibfnamefont {S.}~\bibnamefont {Rotter}},\ }\href {\doibase
  10.1103/PhysRevE.82.021129} {\bibfield  {journal} {\bibinfo  {journal} {Phys
  Rev E}\ }\textbf {\bibinfo {volume} {82}},\ \bibinfo {pages} {021129}
  (\bibinfo {year} {2010})}\BibitemShut {NoStop}%
\bibitem [{\citenamefont {Truccolo}\ \emph {et~al.}(2010)\citenamefont
  {Truccolo}, \citenamefont {Hochberg},\ and\ \citenamefont
  {Donoghue}}]{Truccolo2010a}%
  \BibitemOpen
  \bibfield  {author} {\bibinfo {author} {\bibfnamefont {W.}~\bibnamefont
  {Truccolo}}, \bibinfo {author} {\bibfnamefont {L.~R.}\ \bibnamefont
  {Hochberg}}, \ and\ \bibinfo {author} {\bibfnamefont {J.~P.}\ \bibnamefont
  {Donoghue}},\ }\href {\doibase 10.1038/nn.2455} {\bibfield  {journal}
  {\bibinfo  {journal} {Nat Neurosci}\ }\textbf {\bibinfo {volume} {13}},\
  \bibinfo {pages} {105} (\bibinfo {year} {2010})}\BibitemShut {NoStop}%
\bibitem [{\citenamefont {Meyer}\ and\ \citenamefont {van
  Vreeswijk}(2002)}]{Meyer2002}%
  \BibitemOpen
  \bibfield  {author} {\bibinfo {author} {\bibfnamefont {C.}~\bibnamefont
  {Meyer}}\ and\ \bibinfo {author} {\bibfnamefont {C.}~\bibnamefont {van
  Vreeswijk}},\ }\href {\doibase 10.1162/08997660252741167} {\bibfield
  {journal} {\bibinfo  {journal} {Neural Comput}\ }\textbf {\bibinfo {volume}
  {14}},\ \bibinfo {pages} {369} (\bibinfo {year} {2002})}\BibitemShut
  {NoStop}%
\bibitem [{\citenamefont {Liu}\ and\ \citenamefont {Wang}(2001)}]{Liu2001}%
  \BibitemOpen
  \bibfield  {author} {\bibinfo {author} {\bibfnamefont {Y.~H.}\ \bibnamefont
  {Liu}}\ and\ \bibinfo {author} {\bibfnamefont {X.~J.}\ \bibnamefont {Wang}},\
  }\href {\doibase 10.1023/A:1008916026143} {\bibfield  {journal} {\bibinfo
  {journal} {J Comput Neurosci}\ }\textbf {\bibinfo {volume} {10}},\ \bibinfo
  {pages} {25} (\bibinfo {year} {2001})}\BibitemShut {NoStop}%
\bibitem [{\citenamefont {Jolivet}\ \emph {et~al.}(2006)\citenamefont
  {Jolivet}, \citenamefont {Rauch}, \citenamefont {L\"uscher},\ and\
  \citenamefont {Gerstner}}]{Jolivet2006}%
  \BibitemOpen
  \bibfield  {author} {\bibinfo {author} {\bibfnamefont {R.}~\bibnamefont
  {Jolivet}}, \bibinfo {author} {\bibfnamefont {A.}~\bibnamefont {Rauch}},
  \bibinfo {author} {\bibfnamefont {H.-R.}\ \bibnamefont {L\"uscher}}, \ and\
  \bibinfo {author} {\bibfnamefont {W.}~\bibnamefont {Gerstner}},\ }\href
  {\doibase 10.1007/s10827-006-7074-5} {\bibfield  {journal} {\bibinfo
  {journal} {J Comput Neurosci}\ }\textbf {\bibinfo {volume} {21}},\ \bibinfo
  {pages} {35} (\bibinfo {year} {2006})}\BibitemShut {NoStop}%
\bibitem [{\citenamefont {Zohary}\ \emph {et~al.}(1994)\citenamefont {Zohary},
  \citenamefont {Shadlen},\ and\ \citenamefont {Newsome}}]{Zohary1994}%
  \BibitemOpen
  \bibfield  {author} {\bibinfo {author} {\bibfnamefont {E.}~\bibnamefont
  {Zohary}}, \bibinfo {author} {\bibfnamefont {M.~N.}\ \bibnamefont {Shadlen}},
  \ and\ \bibinfo {author} {\bibfnamefont {W.~T.}\ \bibnamefont {Newsome}},\
  }\href {\doibase 10.1038/370140a0} {\bibfield  {journal} {\bibinfo  {journal}
  {Nature}\ }\textbf {\bibinfo {volume} {370}},\ \bibinfo {pages} {140}
  (\bibinfo {year} {1994})}\BibitemShut {NoStop}%
\bibitem [{\citenamefont {Mar}\ \emph {et~al.}(1999)\citenamefont {Mar},
  \citenamefont {Chow}, \citenamefont {Gerstner}, \citenamefont {Adams},\ and\
  \citenamefont {Collins}}]{Mar1999}%
  \BibitemOpen
  \bibfield  {author} {\bibinfo {author} {\bibfnamefont {D.~J.}\ \bibnamefont
  {Mar}}, \bibinfo {author} {\bibfnamefont {C.~C.}\ \bibnamefont {Chow}},
  \bibinfo {author} {\bibfnamefont {W.}~\bibnamefont {Gerstner}}, \bibinfo
  {author} {\bibfnamefont {R.~W.}\ \bibnamefont {Adams}}, \ and\ \bibinfo
  {author} {\bibfnamefont {J.~J.}\ \bibnamefont {Collins}},\ }\href {\doibase
  10.1073/pnas.96.18.10450} {\bibfield  {journal} {\bibinfo  {journal} {Proc
  Natl Acad Sci U S A}\ }\textbf {\bibinfo {volume} {96}},\ \bibinfo {pages}
  {10450} (\bibinfo {year} {1999})}\BibitemShut {NoStop}%
\bibitem [{\citenamefont {Sompolinsky}\ \emph {et~al.}(2001)\citenamefont
  {Sompolinsky}, \citenamefont {Yoon}, \citenamefont {Kang},\ and\
  \citenamefont {Shamir}}]{Sompolinsky2001}%
  \BibitemOpen
  \bibfield  {author} {\bibinfo {author} {\bibfnamefont {H.}~\bibnamefont
  {Sompolinsky}}, \bibinfo {author} {\bibfnamefont {H.}~\bibnamefont {Yoon}},
  \bibinfo {author} {\bibfnamefont {K.}~\bibnamefont {Kang}}, \ and\ \bibinfo
  {author} {\bibfnamefont {M.}~\bibnamefont {Shamir}},\ }\href {\doibase
  10.1103/PhysRevE.64.051904} {\bibfield  {journal} {\bibinfo  {journal} {Phys
  Rev E}\ }\textbf {\bibinfo {volume} {64}},\ \bibinfo {pages} {051904}
  (\bibinfo {year} {2001})}\BibitemShut {NoStop}%
\bibitem [{\citenamefont {Wang}(2002)}]{Wang2002a}%
  \BibitemOpen
  \bibfield  {author} {\bibinfo {author} {\bibfnamefont {X.-J.}\ \bibnamefont
  {Wang}},\ }\href {\doibase 10.1016/S0896-6273(02)01092-9} {\bibfield
  {journal} {\bibinfo  {journal} {Neuron}\ }\textbf {\bibinfo {volume} {36}},\
  \bibinfo {pages} {955} (\bibinfo {year} {2002})}\BibitemShut {NoStop}%
\bibitem [{\citenamefont {Deco}\ and\ \citenamefont {Rolls}(2006)}]{Deco2006}%
  \BibitemOpen
  \bibfield  {author} {\bibinfo {author} {\bibfnamefont {G.}~\bibnamefont
  {Deco}}\ and\ \bibinfo {author} {\bibfnamefont {E.~T.}\ \bibnamefont
  {Rolls}},\ }\href {\doibase 10.1111/j.1460-9568.2006.04940.x} {\bibfield
  {journal} {\bibinfo  {journal} {Eur J Neurosci}\ }\textbf {\bibinfo {volume}
  {24}},\ \bibinfo {pages} {901} (\bibinfo {year} {2006})}\BibitemShut
  {NoStop}%
\bibitem [{\citenamefont {Renart}\ \emph {et~al.}(2010)\citenamefont {Renart},
  \citenamefont {de~la Rocha}, \citenamefont {Bartho}, \citenamefont
  {Hollender}, \citenamefont {Parga}, \citenamefont {Reyes},\ and\
  \citenamefont {Harris}}]{Renart2010}%
  \BibitemOpen
  \bibfield  {author} {\bibinfo {author} {\bibfnamefont {A.}~\bibnamefont
  {Renart}}, \bibinfo {author} {\bibfnamefont {J.}~\bibnamefont {de~la Rocha}},
  \bibinfo {author} {\bibfnamefont {P.}~\bibnamefont {Bartho}}, \bibinfo
  {author} {\bibfnamefont {L.}~\bibnamefont {Hollender}}, \bibinfo {author}
  {\bibfnamefont {N.}~\bibnamefont {Parga}}, \bibinfo {author} {\bibfnamefont
  {A.}~\bibnamefont {Reyes}}, \ and\ \bibinfo {author} {\bibfnamefont {K.~D.}\
  \bibnamefont {Harris}},\ }\href {\doibase 10.1126/science.1179850} {\bibfield
   {journal} {\bibinfo  {journal} {Science}\ }\textbf {\bibinfo {volume}
  {327}},\ \bibinfo {pages} {587} (\bibinfo {year} {2010})}\BibitemShut
  {NoStop}%
\bibitem [{\citenamefont {Stratonovich}(1963)}]{Stratonovich1963}%
  \BibitemOpen
  \bibfield  {author} {\bibinfo {author} {\bibfnamefont {R.~L.}\ \bibnamefont
  {Stratonovich}},\ }\href@noop {} {\emph {\bibinfo {title} {Topics in the
  theory of random noise I}}},\ Mathematics and its applications; vol. 3\
  (\bibinfo  {publisher} {New York: Gordon and Breach},\ \bibinfo {year}
  {1963})\BibitemShut {NoStop}%
\bibitem [{\citenamefont {Shannon}(1948)}]{Shannon1948}%
  \BibitemOpen
  \bibfield  {author} {\bibinfo {author} {\bibfnamefont {C.~E.}\ \bibnamefont
  {Shannon}},\ }\href {\doibase 10.1002/j.1538-7305.1948.tb01338.x} {\bibfield
  {journal} {\bibinfo  {journal} {Bell Syst. Tech. J.}\ }\textbf {\bibinfo
  {volume} {27}},\ \bibinfo {pages} {379} (\bibinfo {year} {1948})}\BibitemShut
  {NoStop}%
\bibitem [{\citenamefont {Gabbiani}(1996)}]{Gabbiani1996}%
  \BibitemOpen
  \bibfield  {author} {\bibinfo {author} {\bibfnamefont {F.}~\bibnamefont
  {Gabbiani}},\ }\href {\doibase 10.1088/0954-898X/7/1/005} {\bibfield
  {journal} {\bibinfo  {journal} {Network Comp Neural}\ }\textbf {\bibinfo
  {volume} {7}},\ \bibinfo {pages} {61} (\bibinfo {year} {1996})}\BibitemShut
  {NoStop}%
\bibitem [{\citenamefont {Borst}\ and\ \citenamefont
  {Theunissen}(1999)}]{Borst1999}%
  \BibitemOpen
  \bibfield  {author} {\bibinfo {author} {\bibfnamefont {A.}~\bibnamefont
  {Borst}}\ and\ \bibinfo {author} {\bibfnamefont {F.~E.}\ \bibnamefont
  {Theunissen}},\ }\href {\doibase 10.1038/14731} {\bibfield  {journal}
  {\bibinfo  {journal} {Nat Neurosci}\ }\textbf {\bibinfo {volume} {2}},\
  \bibinfo {pages} {947} (\bibinfo {year} {1999})}\BibitemShut {NoStop}%
\bibitem [{\citenamefont {Chacron}\ \emph {et~al.}(2004)\citenamefont
  {Chacron}, \citenamefont {Lindner},\ and\ \citenamefont
  {Longtin}}]{Chacron2004}%
  \BibitemOpen
  \bibfield  {author} {\bibinfo {author} {\bibfnamefont {M.~J.}\ \bibnamefont
  {Chacron}}, \bibinfo {author} {\bibfnamefont {B.}~\bibnamefont {Lindner}}, \
  and\ \bibinfo {author} {\bibfnamefont {A.}~\bibnamefont {Longtin}},\ }\href
  {\doibase 10.1103/PhysRevLett.92.080601} {\bibfield  {journal} {\bibinfo
  {journal} {Phys Rev Lett}\ }\textbf {\bibinfo {volume} {92}},\ \bibinfo
  {pages} {080601} (\bibinfo {year} {2004})}\BibitemShut {NoStop}%
\bibitem [{\citenamefont {Vilela}\ and\ \citenamefont
  {Lindner}(2009)}]{Vilela2009}%
  \BibitemOpen
  \bibfield  {author} {\bibinfo {author} {\bibfnamefont {R.~D.}\ \bibnamefont
  {Vilela}}\ and\ \bibinfo {author} {\bibfnamefont {B.}~\bibnamefont
  {Lindner}},\ }\href {\doibase 10.1103/PhysRevE.80.031909} {\bibfield
  {journal} {\bibinfo  {journal} {Phys Rev E}\ }\textbf {\bibinfo {volume}
  {80}},\ \bibinfo {pages} {031909} (\bibinfo {year} {2009})}\BibitemShut
  {NoStop}%
\bibitem [{\citenamefont {Chacron}\ \emph {et~al.}(2005)\citenamefont
  {Chacron}, \citenamefont {Maler},\ and\ \citenamefont
  {Bastian}}]{Chacron2005b}%
  \BibitemOpen
  \bibfield  {author} {\bibinfo {author} {\bibfnamefont {M.~J.}\ \bibnamefont
  {Chacron}}, \bibinfo {author} {\bibfnamefont {L.}~\bibnamefont {Maler}}, \
  and\ \bibinfo {author} {\bibfnamefont {J.}~\bibnamefont {Bastian}},\ }\href
  {\doibase 10.1038/nn1433} {\bibfield  {journal} {\bibinfo  {journal} {Nat
  Neurosci}\ }\textbf {\bibinfo {volume} {8}},\ \bibinfo {pages} {673}
  (\bibinfo {year} {2005})}\BibitemShut {NoStop}%
\bibitem [{\citenamefont {Krahe}\ \emph {et~al.}(2008)\citenamefont {Krahe},
  \citenamefont {Bastian},\ and\ \citenamefont {Chacron}}]{Krahe2008}%
  \BibitemOpen
  \bibfield  {author} {\bibinfo {author} {\bibfnamefont {R.}~\bibnamefont
  {Krahe}}, \bibinfo {author} {\bibfnamefont {J.}~\bibnamefont {Bastian}}, \
  and\ \bibinfo {author} {\bibfnamefont {M.~J.}\ \bibnamefont {Chacron}},\
  }\href {\doibase 10.1152/jn.90300.2008} {\bibfield  {journal} {\bibinfo
  {journal} {J Neurophysiol}\ }\textbf {\bibinfo {volume} {100}},\ \bibinfo
  {pages} {852} (\bibinfo {year} {2008})}\BibitemShut {NoStop}%
\bibitem [{\citenamefont {Richardson}(2009)}]{Richardson2009}%
  \BibitemOpen
  \bibfield  {author} {\bibinfo {author} {\bibfnamefont {M.~J.~E.}\
  \bibnamefont {Richardson}},\ }\href {\doibase 10.1103/PhysRevE.80.021928}
  {\bibfield  {journal} {\bibinfo  {journal} {Phys Rev E}\ }\textbf {\bibinfo
  {volume} {80}},\ \bibinfo {pages} {021928} (\bibinfo {year}
  {2009})}\BibitemShut {NoStop}%
\bibitem [{\citenamefont {Rosenbaum}\ \emph {et~al.}(2012)\citenamefont
  {Rosenbaum}, \citenamefont {Rubin},\ and\ \citenamefont
  {Doiron}}]{Rosenbaum2012}%
  \BibitemOpen
  \bibfield  {author} {\bibinfo {author} {\bibfnamefont {R.}~\bibnamefont
  {Rosenbaum}}, \bibinfo {author} {\bibfnamefont {J.}~\bibnamefont {Rubin}}, \
  and\ \bibinfo {author} {\bibfnamefont {B.}~\bibnamefont {Doiron}},\ }\href
  {\doibase 10.1371/journal.pcbi.1002557} {\bibfield  {journal} {\bibinfo
  {journal} {PLoS Comput Biol}\ }\textbf {\bibinfo {volume} {8}},\ \bibinfo
  {pages} {e1002557} (\bibinfo {year} {2012})}\BibitemShut {NoStop}%
\bibitem [{\citenamefont {Sharafi}\ \emph {et~al.}(2013)\citenamefont
  {Sharafi}, \citenamefont {Benda},\ and\ \citenamefont
  {Lindner}}]{Sharafi2013}%
  \BibitemOpen
  \bibfield  {author} {\bibinfo {author} {\bibfnamefont {N.}~\bibnamefont
  {Sharafi}}, \bibinfo {author} {\bibfnamefont {J.}~\bibnamefont {Benda}}, \
  and\ \bibinfo {author} {\bibfnamefont {B.}~\bibnamefont {Lindner}},\ }\href
  {\doibase 10.1007/s10827-012-0421-9} {\bibfield  {journal} {\bibinfo
  {journal} {J Comput Neurosci}\ }\textbf {\bibinfo {volume} {34}},\ \bibinfo
  {pages} {285} (\bibinfo {year} {2013})}\BibitemShut {NoStop}%
\bibitem [{\citenamefont {Droste}\ \emph {et~al.}(2013)\citenamefont {Droste},
  \citenamefont {Schwalger},\ and\ \citenamefont {Lindner}}]{Droste2013}%
  \BibitemOpen
  \bibfield  {author} {\bibinfo {author} {\bibfnamefont {F.}~\bibnamefont
  {Droste}}, \bibinfo {author} {\bibfnamefont {T.}~\bibnamefont {Schwalger}}, \
  and\ \bibinfo {author} {\bibfnamefont {B.}~\bibnamefont {Lindner}},\ }\href
  {\doibase 10.3389/fncom.2013.00086} {\bibfield  {journal} {\bibinfo
  {journal} {Front Comput Neurosci}\ }\textbf {\bibinfo {volume} {7}},\
  \bibinfo {pages} {86} (\bibinfo {year} {2013})}\BibitemShut {NoStop}%
\bibitem [{\citenamefont {El~Hady}\ \emph {et~al.}(2013)\citenamefont
  {El~Hady}, \citenamefont {Afshar}, \citenamefont {Br\"oking}, \citenamefont
  {Schl\"uter}, \citenamefont {Geisel}, \citenamefont {St\"uhmer},\ and\
  \citenamefont {Wolf}}]{ElHady2013}%
  \BibitemOpen
  \bibfield  {author} {\bibinfo {author} {\bibfnamefont {A.}~\bibnamefont
  {El~Hady}}, \bibinfo {author} {\bibfnamefont {G.}~\bibnamefont {Afshar}},
  \bibinfo {author} {\bibfnamefont {K.}~\bibnamefont {Br\"oking}}, \bibinfo
  {author} {\bibfnamefont {O.~M.}\ \bibnamefont {Schl\"uter}}, \bibinfo
  {author} {\bibfnamefont {T.}~\bibnamefont {Geisel}}, \bibinfo {author}
  {\bibfnamefont {W.}~\bibnamefont {St\"uhmer}}, \ and\ \bibinfo {author}
  {\bibfnamefont {F.}~\bibnamefont {Wolf}},\ }\href {\doibase
  10.3389/fncir.2013.00167} {\bibfield  {journal} {\bibinfo  {journal} {Front
  Neural Circuits}\ }\textbf {\bibinfo {volume} {7}},\ \bibinfo {pages} {167}
  (\bibinfo {year} {2013})}\BibitemShut {NoStop}%
\bibitem [{\citenamefont {Gewaltig}\ and\ \citenamefont
  {Diesmann}(2007)}]{Gewaltig:NEST}%
  \BibitemOpen
  \bibfield  {author} {\bibinfo {author} {\bibfnamefont {M.-O.}\ \bibnamefont
  {Gewaltig}}\ and\ \bibinfo {author} {\bibfnamefont {M.}~\bibnamefont
  {Diesmann}},\ }\href {\doibase 10.4249/scholarpedia.1430} {\bibfield
  {journal} {\bibinfo  {journal} {Scholarpedia}\ }\textbf {\bibinfo {volume}
  {2}},\ \bibinfo {pages} {1430} (\bibinfo {year} {2007})}\BibitemShut
  {NoStop}%
\end{thebibliography}%

\end{document}